\documentclass[journal]{IEEEtran}
\usepackage{amsmath,amsfonts}
\usepackage{algorithmic}
\usepackage{algorithm}
\usepackage{array}
\usepackage[caption=false,font=normalsize,labelfont=sf,textfont=sf]{subfig}
\usepackage{textcomp}
\usepackage{stfloats}
\usepackage{url}
\usepackage{verbatim}
\usepackage{graphicx}
\usepackage{bm}
\usepackage{multirow}
\usepackage{multicol}
\usepackage{booktabs}
\usepackage{hhline}
\usepackage{graphicx} 
\usepackage{orcidlink}
\usepackage{cite}
\usepackage{xcolor}

\def\rts{\textit{RT}$_{60}$}

\begin{document}

\title{DARAS: Dynamic Audio-Room Acoustic Synthesis\\ for Blind Room Impulse Response Estimation}

\author{Chunxi Wang~\orcidlink{0009-0009-0625-3934},~\IEEEmembership{Student Member,~IEEE,}~Maoshen Jia~\orcidlink{0000-0002-3452-3913},~\IEEEmembership{Senior Member,~IEEE,} and~Wenyu Jin,~\IEEEmembership{Member,~IEEE}
\vspace{-8mm}
\thanks{This work was supported in part by the National Natural Science Foundation of China under Grant 62471012, in part by Beijing Natural Science Foundation under Grant L233032 and Grant L223033. \textit{(Corresponding author: Maoshen Jia.) }

Chunxi Wang and Maoshen Jia are with the School of Information Science and Technology, Beijing University of Technology, Beijing 100124, China (e-mail: chunxiwang@emails.bjut.edu.cn; jiamaoshen@bjut.edu.cn). 

Wenyu Jin is with the Unseen AI Inc., New York 11201, USA (e-mail:
wenyu@unseen-ai.com).
\vspace{-2mm}
\begin{flushleft}
Audio samples are available at: \url{https://github.com/bjut-chunxiwang/DARAS}
\end{flushleft}

}}

\markboth{Journal of \LaTeX\ Class Files,~Vol.~14, No.~8, August~2021}%
{Shell \MakeLowercase{\textit{et al.}}: A Sample Article Using IEEEtran.cls for IEEE Journals}


\maketitle

\begin{abstract}
Room Impulse Responses (RIRs) accurately characterize acoustic properties of indoor environments and play a crucial role in applications such as speech enhancement, speech recognition, and audio rendering in augmented reality (AR) and virtual reality (VR). Existing blind estimation methods struggle to achieve practical accuracy. To overcome this challenge, we propose the dynamic audio-room acoustic synthesis (DARAS) model, a novel deep learning framework that is explicitly designed for blind RIR estimation from monaural reverberant speech signals. First, a dedicated deep audio encoder effectively extracts relevant nonlinear latent space features. Second, the Mamba-based self-supervised blind room parameter estimation (MASS-BRPE) module, utilizing the efficient Mamba state space model (SSM), accurately estimates key room acoustic parameters and features. Third, the system incorporates a hybrid-path cross-attention feature fusion module, enhancing deep integration between audio and room acoustic features. Finally, our proposed dynamic acoustic tuning (DAT) decoder adaptively segments early reflections and late reverberation to improve the realism of synthesized RIRs. Experimental results, including a MUSHRA-based subjective listening study, demonstrate that DARAS substantially outperforms existing baseline models, providing a robust and effective solution for practical blind RIR estimation in real-world acoustic environments.
\end{abstract}


\vspace{-2mm}
\section{Introduction}
\IEEEPARstart{T}{he}  Room Impulse Response (RIR) is a core parameter that describes acoustic characteristics of a room. It comprehensively reflects the processes of sound wave propagation, reflection, refraction, and attenuation within the indoor environment. When modeling room acoustics as a linear time-invariant system, the RIR precisely characterizes the sound wave propagation path from the source to the receiver, revealing the acoustic properties of the room. An accurate estimation of the RIR is crucial for analyzing and understanding the room’s acoustic environment, as it showcases the interaction between sound waves and the objects within the room.

\vspace{-4mm}
\subsection{Literature Review}
Although RIRs have widespread applications in speech enhancement \cite{gannot2017consolidated,Lemercier2025unsupervised}, speech recognition \cite{Zhang2018,wang2025vinp}, room acoustics modeling \cite{lehmann2010diffuse,luo2024fast,jin2022acoustic} and sound field rendering \cite{betlehem2005theory,jin15theory, wang2024hearing}, their direct measurements still face many challenges. Traditional methods rely on specific test signals, require a quiet environment, and use high-quality audio equipment. However, background noise, environmental interference and equipment limitations often affect measurement accuracy \cite{szoke2019building}. Moreover, the absorption characteristics and reverberation effects of a room can vary in real-world environments due to changes in activities and layouts \cite{choi2018effects}. As an alternative, acoustic simulators \cite{shuku1973analysis, allen1979image, krokstad1968calculating} generate RIRs by simulating sound wave propagation. They take into account factors such as room geometry, positions of sound sources and microphones, and reflection characteristics. However, these methods are computationally complex, and their simulation accuracy is limited when dealing with rooms of intricate shapes.

To overcome the above-mentioned challenges, recent studies have shifted towards blind estimation methods for key room acoustic parameters, such as room volume \cite{yu2020room, ick2023blind, srivastava2021blind, wang2024attention, wang2024berp, wang2024exploring}, reverberation time (\rts) \cite{xiong2018exploring, deng2020online, gotz2022blind, duangpummet2022blind, eaton2016estimation, saini2023blind, wang2025ss}, and direct-to-reverberant ratio (DRR) \cite{eaton2016estimation, bryan2020impulse}.  These methods do not require direct RIR measurement. Instead, they estimate the room's acoustic characteristics by collecting signals using commercial microphones and other consumer devices.

However, in applications requiring high precision and fidelity, such as augmented reality (AR), virtual reality (VR), and reverb matching in audio post-production, limitations of blind estimation methods become increasingly apparent. In these scenarios, a single parameter or a finite set of parameters cannot comprehensively describe acoustic characteristics of the room, making it difficult to meet accuracy requirements \cite{peters2012matching, sarroff2020blind, koo2021reverb}. 

Compared to estimating discrete acoustic parameters and subsequently synthesizing RIRs, directly estimating RIRs from recorded signals is more practically appealing. RIRs not only provide a qualitative description of key room acoustic parameters, but also reveal the time-frequency characteristics of the sound field. In recent years, extensive work has been reported in directly estimating RIRs from reverberant speech signals. Several traditional signal processing techniques have been proposed \cite{karjalainen2002estimation, crammer2006room, lin2006bayesian, crocco2015room, fu2022sparse}, but these methods often rely on certain assumptions, such as performance dependencies on the choice of regularization parameters \cite{crocco2015room}, assuming the signal source to be a modulated Gaussian pulse rather than real speech signals \cite{karjalainen2002estimation, crocco2015room, fu2022sparse}, or requiring prior knowledge of characteristics of the microphones and speakers \cite{crammer2006room, lin2006bayesian}.  Although these assumptions have demonstrated good performance under controlled conditions, their applicability remains somewhat limited in more complex and unpredictable real-world scenarios.


With continuous advancements in deep learning technology, data-driven RIR estimation models have demonstrated significant advantages. These methods achieve accurate estimation of RIRs by extracting core features from monaural reverberant speech. Considering unique characteristics of RIRs, such as sparsity and sharp transients, some researchers have proposed adaptive generative architectures, including generative adversarial networks (GANs) and the regularized quantized variational autoencoder (RQ-VAE) \cite{ratnarajah2023towards, lee2023yet, liao2023blind}. However, these methods typically perform better at estimating early RIR components (i.e., direct sound and early reflections) and are more vulnerable when estimating late reverberation components. This is because early RIR components exhibit sparse, impulsive characteristics, whereas late components exhibit noise-like structures with significantly lower amplitude.

Inspired by properties of room acoustics, the filtered noise shaping (FiNS) network \cite{steinmetz2021filtered} provides an efficient estimation approach by modeling RIRs as a combination of direct sound, early reflections, and decaying filtered noise signals through a time-domain encoder-decoder architecture. Additionally, the Demucs-based blind RIR estimation (Dbre) model \cite{yapar2024demucs, lu2025rir} adopts an encoder-decoder architecture with U-Net skip connections, further enhancing the accuracy of RIR estimation.
\vspace{-8mm}
\subsection{Approach and Contribution by This Paper}
Although the above-mentioned data-driven RIR estimation models have made significant progress to some extent, they still face notable challenges and limitations. Existing models typically rely on audio features extracted from reverberant speech for RIR estimation. In real-world scenarios, geometries, structural features, and acoustic properties of rooms are often complex and variable \cite{ratnarajah2022mesh2ir,ratnarajah2024av}, especially in spaces with irregular layouts or diverse building materials. Physical and acoustic characteristics of these rooms are typically reflected as complex components in the reverberant signal. Existing estimation models may struggle to accurately separate key acoustic components in these circumstances, such as direct sound, early reflections, and late reflections.

Therefore, inadequate modeling of room parameters may result in the model’s insufficient understanding of the room’s acoustic characteristics, causing the generated RIRs to fail to accurately and faithfully reflect the actual room parameters. These discrepancies not only affect the accuracy of acoustic models but may also have detrimental impacts on subsequent sound processing tasks.

To address challenges in blind RIR estimation such as difficulties in data collection, high annotation costs, and mismatches between estimated RIRs and actual room acoustic parameters, this paper proposes a novel model, referred to as the \emph{dynamic audio-room acoustic synthesis} (DARAS) blind RIR estimation model. The DARAS model estimates audio and room acoustic features from reverberant speech signals and adaptively adjusts the generated RIRs through the deep fusion of these features, achieving greater accuracy in matching real acoustic environments.
The main contributions of this paper are as follows:

\emph{1) Deep audio encoder:} A specialized deep convolutional neural network-based audio encoder is designed explicitly for blind RIR estimation tasks. It extracts nonlinear latent space features from reverberant speech. Compared to traditional shallow feature extraction methods, this deep audio encoder significantly enhances the representational capacity of audio features.

\emph{2) Mamba-based self-supervised blind room parameter estimation (MASS-BRPE):} A novel room parameter estimation module based on the Mamba state space model (SSM), named MASS-BRPE, is proposed. It can blindly estimate critical room acoustic parameters along with the corresponding room acoustic features. Compared to previous Transformer-based models \cite{wang2025ss}, MASS-BRPE maintains comparable estimation accuracy while significantly reducing GPU memory usage and computational inference time, greatly enhancing the efficiency and scalability of the model in practical RIR estimation tasks.

\emph{3) Hybrid-path cross-attention feature fusion:} A hybrid-path cross-attention feature fusion module is proposed, aiming to capture the complex and dynamic dependencies between audio and room acoustic features. This module employs a dual-path parallel interactive architecture to facilitate multi-level feature fusion, effectively capturing realistic and fine-grained dynamic correlations between acoustic and audio features. Experimental results demonstrate that this module notably improves the effectiveness of the feature fusion, facilitating the adaptability of RIR estimation across diverse complex acoustic scenarios.

\emph{4) Dynamic acoustic tuning (DAT) Decoder:} A decoder capable of adaptively distinguishing early reflections from late reverberation is designed. Utilizing dynamically estimated boundary points, this decoder segments RIRs into early and late stages, separately modeling and synthesizing each segment. This approach significantly improves the accuracy of the generated RIRs by reproducing authentic room acoustic characteristics.

\emph{5) Multiple stimuli with hidden reference and anchor (MUSHRA) subjective listening study:} A MUSHRA test \cite{series2014method} was conducted to evaluate the perceptual similarity of reverberant audio generated by different models. The results show that RIRs produced by the proposed DARAS model are perceptually closer to real-world recordings and outperform baseline methods, confirming the effectiveness of the proposed method.

The DARAS model effectively mitigates challenges such as difficulties in data acquisition and high annotation costs in RIR estimation tasks, while enabling precise blind RIR estimation. Experimental results demonstrate that through dynamic fusion of estimated audio and room acoustic features, the DARAS model exhibits good generalization and estimation accuracy across various complex acoustic environments, which leads to an accurate match between generated RIRs and actual room conditions, thereby better fulfilling real-world auditory experience requirements.

The rest of this paper is organized as follows. In Section \ref{METHODOLOGY}, we present the overall architecture of the proposed DARAS model. Section \ref{Experimental Procedure} describes  the experimental procedure and the construction of the dataset. Experimental results based on previously unseen real-world rooms, focusing on the RIR estimation accuracy of different models and their perceived audio quality in actual environments are presented in Section \ref{EXPERIMENTAL RESULTS}. Finally, Section \ref{CONCLUSION} provides the conclusion.

\vspace{-2mm}
\section{METHODOLOGY}\label{METHODOLOGY}

\begin{figure}[ht]
\centering
\includegraphics[width=8.6cm]{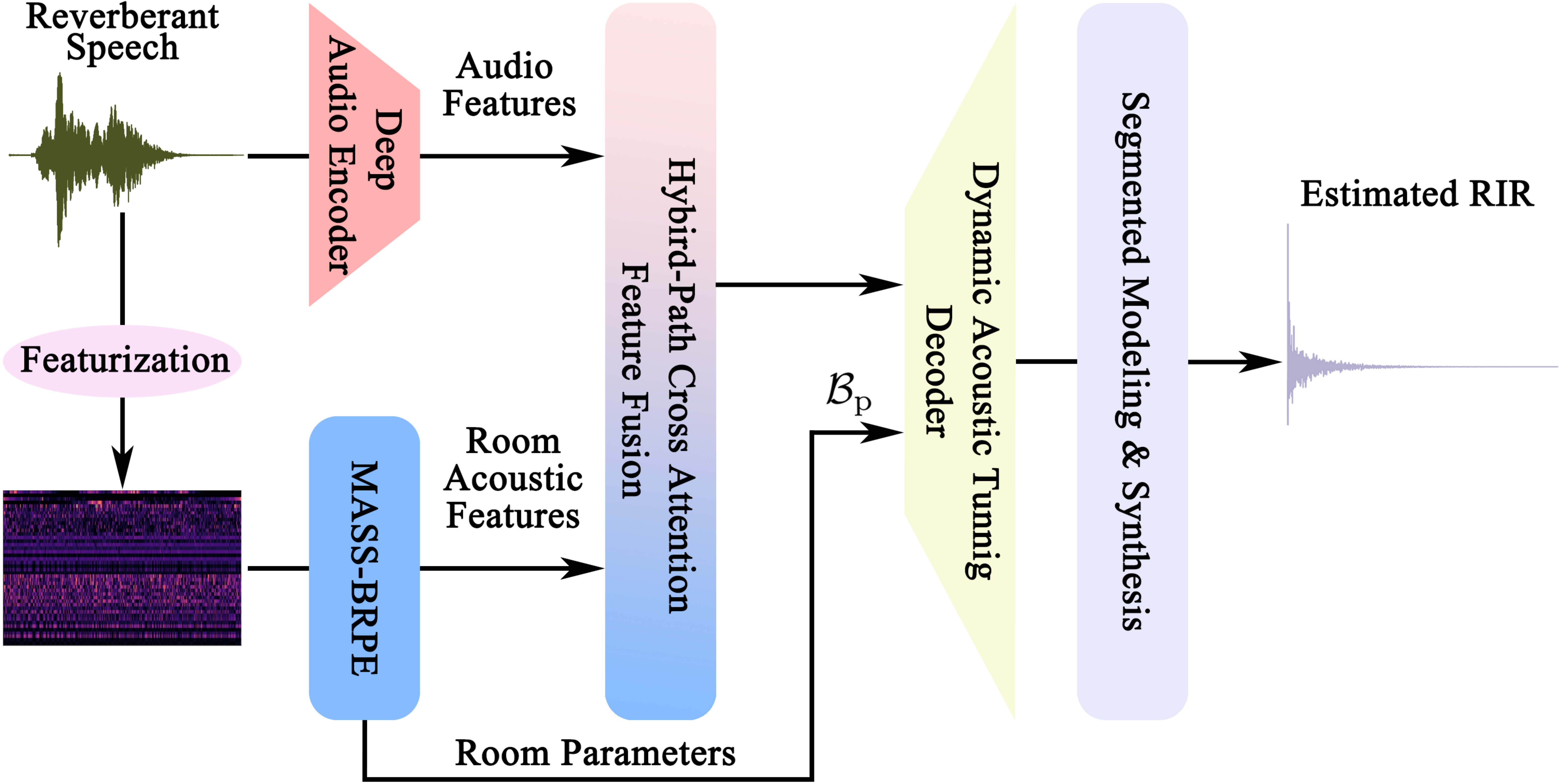}
\vspace{-3mm}
\caption{Overview of the DARAS Blind RIR Estimation Model. 
This figure illustrates the proposed DARAS model designed to estimate RIR from monaural reverberant speech. The model comprises four modules: (1) a Deep Audio Encoder extracting nonlinear features from reverberant speech; (2) the MASS-BRPE module, employing state space models (SSMs) to estimate room acoustic parameters and features; (3) a Hybrid-Path Cross-attention Feature Fusion module, dynamically guiding audio features integration with room acoustic features to achieve refined reverberation-aware representations; and (4) a DAT Decoder, adaptively segmenting RIR into early reflections and late reverberation stages based on the boundary point ($\mathcal{B}_\mathrm{p}$) estimated by the MASS-BRPE module, synthesizing each stage individually.
}
\vspace{-3mm}
\label{fig:od}
\end{figure}

\begin{figure}[ht]
\centering
\includegraphics[width=8.5cm]{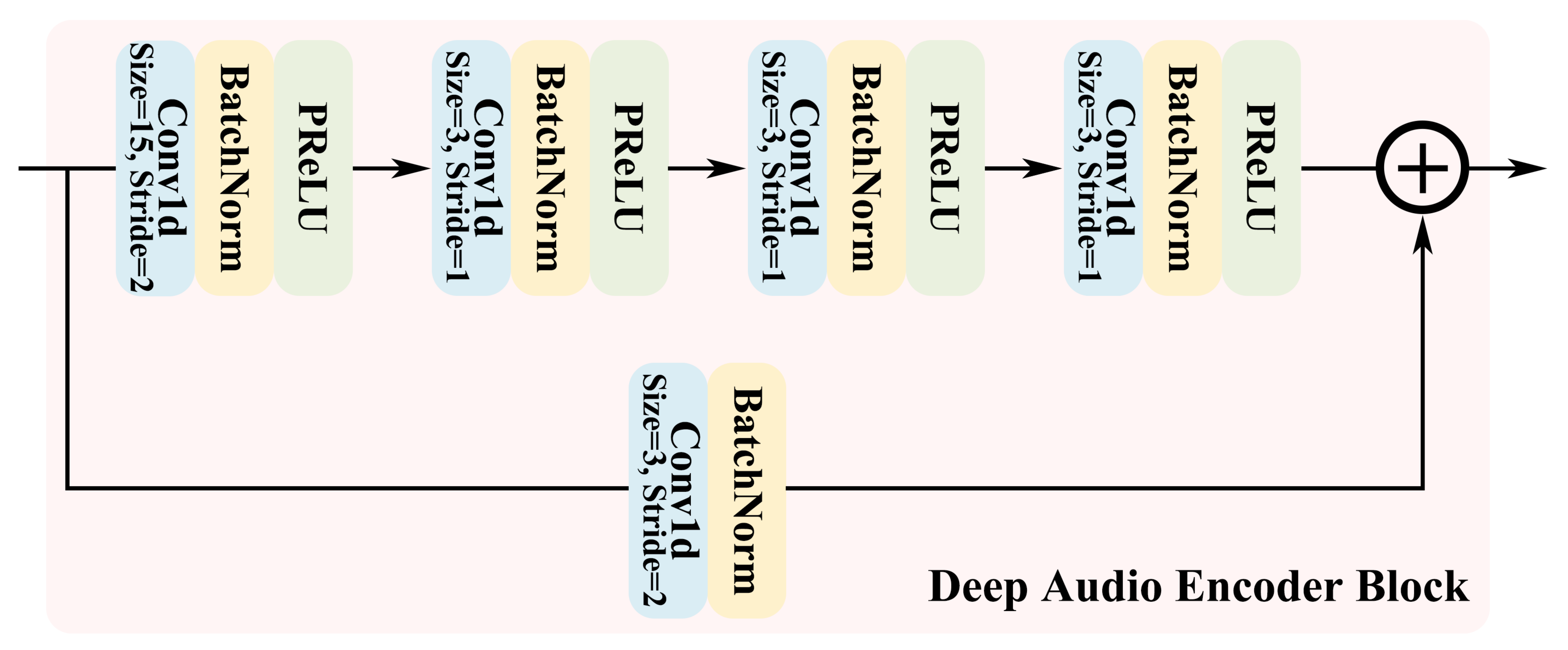}
\vspace{-3mm}
\caption{The architecture of the deep audio encoder block.}
\vspace{-6mm}
\label{fig:daed}
\end{figure}

Reverberant speech $\bm{\mathbf{s}}_{\mathrm{r}}(n)$ is formulated as the convolution of anechoic speech $\bm{\mathbf{s}}_{\mathrm{a}}(n)\in \mathbb{R}^{1 \times T}$ with RIR $\bm{\mathbf{h}}(n)\in \mathbb{R}^{1 \times U}$, with additive noise $\mathbf{w}(n)$, as follows:
\begin{equation}
     \bm{\mathbf{s}}_{\mathrm{r}}(n)=\bm{\mathbf{s}}_{\mathrm{a}}(n) * \bm{\mathbf{h}}(n)+\mathbf{w}(n),
\end{equation}
where $*$ denotes the convolution operator, $n$ is the discrete-time index, and $T$ and $U$ respectively represent the total number of samples in the anechoic speech and the RIR.

This section presents the methodology of our proposed DARAS model, which aims to estimate the RIR from monaural reverberant speech $\bm{\mathbf{s}}_{\mathrm{r}}(n)$. The overview of the proposed model is shown in Fig. \ref{fig:od}. Our approach consists of extracting comprehensive deep audio and room acoustic features, integrating these features using a novel hybrid-path cross-attention fusion module, and dynamically segmenting the RIR components to more accurately reflect real-world acoustic environments.

\vspace{-4mm}
\subsection{Deep Audio Encoder}

A deep audio encoder is proposed to enhance the representational capacity of audio features through nonlinear feature extraction. The encoder extracts RIR-related features from monaural reverberant speech and maps them to a suitably dimensioned feature space by progressive downsampling. 

Existing blind RIR estimation algorithms have paid limited attention to the encoder, typically using only shallow operators for feature extraction, which may limit the expressiveness of the model \cite{steinmetz2021filtered}. In this work, we explore the capabilities of the deep audio encoder in reverberant speech $\bm{\mathbf{s}}_{\mathrm{r}}(n)$ transformations, as illustrated in Fig. \ref{fig:daed}. Each deep convolutional block comprises four nonlinear encoding layers, with each layer consisting of a 1-D convolution, batch normalization, and a parametric rectified linear unit (PReLU) activation function \cite{he2015delving}. To enhance the continuous transfer of information and prevent the vanishing of gradients, a stride of 2 is introduced with $1\times1$ convolutions for residual connections, followed by batch normalization. As the network depth increases, the number of channels gradually increases, producing an output of 512 channels in the final layer. Finally, the features are reduced to a 128-dimensional space using $1\times1$ convolutions, forming a speech 2-D representation of the audio features $\bm{\mathbf{F}}_{\mathrm{s}}\in \mathbb{R}^{N_\mathrm{s} \times L}$, where $N_\mathrm{s}$ is the feature dimension, and $L$ is the time dimension.

\begin{figure}[!t]
  \centering
  \includegraphics[width=0.9\linewidth]{Fig/BRPE.pdf}
  \vspace{-4mm}
  \caption{Schematic diagram of the overall architecture of the proposed MASS-BRPE module.}
  \label{brpe}
  \vspace{-5mm}
\end{figure}

\vspace{-4mm}
\subsection{MASS-BRPE}

The goal of MASS-BRPE is to estimate key room parameters from reverberant speech $\bm{\mathbf{s}}_{\mathrm{r}}(n)$ and to obtain room acoustic features related to these parameters, as illustrated in Fig. \ref{brpe}. These room parameters include the room volume $\mathcal{V}$, \rts, and a dynamic boundary point $\mathcal{B}_\mathrm{p}$ between early and late reverberation. Compared to traditional in-situ measurement methods, blind estimation is more convenient for labeling, as it does not rely on intrusive measurements or expert knowledge. 

To enhance efficiency by reducing GPU memory consumption and computational inference time, we adopt the latest SSMs, particularly those based on the Mamba framework \cite{gu2023mamba, shams2024ssamba,wu2025cross}. 
Inspired by the work in \cite{gu2023mamba, shams2024ssamba,wang2025ss}, we propose the MASS-BRPE module. During the pretraining phase, input real-valued spectrograms are divided into non-overlapping patches, with a subset randomly masked. Embeddings of these masked patches are then used as training targets, with the classification head and reconstruction head employed to perform discriminative and generative tasks, respectively. Specifically, in the discriminative task, the MASS-BRPE module aims to correctly identify the masked patches, i.e., to select the correct patch for each masked position from all the masked patches. In the generative task, the module focuses on reconstructing the original content of the masked patches. Audio samples from AudioSet-2M \cite{gemmeke2017audio} and LibriSpeech \cite{panayotov2015librispeech} are mixed for self-supervised pretraining of the MASS-BRPE module. These two datasets abandon the labels and focus solely on the audio components.


In the fine-tuning and inference stages, the input and output dimensions of the model are adjusted to accommodate the MASS-BRPE module. Specifically, magnitude features, phase features, and the first-order derivatives of phase coefficients are extracted from the complex Gammatone spectrogram to serve as the module’s inputs. Prior work has shown that low-frequency effects play a crucial role in room acoustic parameter estimation \cite{srivastava2022realistic}. Therefore, we use a Gammatone ERB filter bank consisting of 20 bands, covering a frequency range from 50 Hz to 2000 Hz to generate the complex Gammatone spectrogram $\mathbf{D}_{\mathrm{gram}}$ of the reverberant speech signal $\mathbf{s}_{\mathrm{r}}(n)$. Subsequently, the magnitude features $\mathfrak{X}_{\text{mag}}$ are obtained by taking the magnitude of the complex Gammatone spectrogram $\mathbf{D}_{\mathrm{gram}}$, represented as:
\vspace{-3mm}
\begin{equation}
   \boldsymbol{\mathfrak{X}}_{\text{mag}} = |\mathbf{D}_{\text{gram}}|.
   \vspace{-3mm}
\end{equation}
The phase features $\boldsymbol{\mathfrak{X}}_{\text{phase}}$ are represented by the phase angle of each time–frequency bin:
\vspace{-3mm}
\begin{equation}
   \boldsymbol{\mathfrak{X}}_{\text{phase}} = \arg(\mathbf{D}_{\text{gram}}),
   \vspace{-3mm}
\end{equation}
where $\arg(\cdot)$ denotes the phase angle of a complex number. Furthermore, the first-order derivatives of the phase coefficients are extracted along the frequency axis. We then aggregate these features along the frequency dimension to construct a 2-D feature block. This feature configuration is consistent with the ``+\textit{Phase}" strategy and has been proven to outperform methods based solely on amplitude spectral features in terms of performance \cite{ick2023blind, wang2025ss}.

Subsequently, the 2-D feature block map is divided into $J$ patches $\mathbf{S}_{[j]}$ of size $16 \times 16$. Each patch is first transformed by a linear projection to obtain a 1-D patch embedding $\mathbf{E}_{[j]}$, which is then combined with a learnable positional embedding $\mathbf{P}_{[j]}$, yielding the combined embedding $\mathbf{E}'_{[j]} = \mathbf{E}_{[j]} + \mathbf{P}_{[j]}$. These combined embeddings are subsequently processed through a Mamba encoder consisting of bidirectional SSMs \cite{liang2025vim}, enabling simultaneous extraction of forward and backward temporal dependencies. As described in \cite{shams2024ssamba}, the Bidirectional Mamba block projects $\mathbf{E}'_{[j]}$ into the intermediate variables $x$ and $z$ through linear transformations, where $x = \text{Linear}_x(\mathbf{E}'_{[j]}) $ and $ z = \text{Linear}_z(\mathbf{E}'_{[j]}) $. Here $x$ is used for feature extraction and $ z $ is used to modulate the forward and backward outputs. Specifically, for each temporal direction (forward and backward), $x$ passes through a SiLU activation and a 1-D convolution, generating direction-specific features. These features are modulated by additional linear transformations and gain controls within the SSM. The forward and backward outputs are then merged and further refined by a nonlinear activation followed by a linear projection, producing the final embedding $\mathbf{O}_{[j]}$. This mechanism allows $z$ to serve as an intermediate representation that modulates the outputs of the SSM blocks in both directions, thus enhancing the encoder's ability to capture complex temporal relationships within the audio data.

Specifically, the module takes 2-D feature blocks as input and outputs predicted core room acoustic features and parameter labels. The output patch embedding sequence $\mathbf{O}_{[j]}$ is used to represent the 2-D  feature blocks. By applying mean pooling, a feature representation at the audio clip level is obtained. In this process, global acoustic features related to $\mathcal{V}$ and \rts\ are respectively denoted as $\mathbf{q}_\mathcal{V}$ and $\mathbf{q}_\zeta$. Ultimately, linear layers are used to predict these room parameters. The module employs mean squared error to optimize the accuracy between logarithmic scale true room parameters and estimated room parameters on a logarithmic scale.

To summarize, the MASS-BRPE module facilitates accurate estimation of important room acoustic parameters, including $\mathcal{V}$, \rts, and $\mathcal{B}_\mathrm{p}$. 

\begin{figure}[ht]
\centering
\includegraphics[width=\linewidth]{Fig/HPCA.pdf}
\vspace{-5mm}
\caption{Hybrid‑path cross‑attention feature fusion module.}
\vspace{-6mm}
\label{fig:hpca}
\end{figure}

\vspace{-4mm}
\subsection{Hybrid-Path Cross-Attention Feature Fusion} \label{sec:hybrid-path-attention}

We propose a hybrid-path cross-attention feature fusion module. This module enables more synergistic integration of features through multi-level interactions between room acoustic features and audio features, thereby generating RIRs for unseen real-world rooms. 

In particular, both room acoustic features and audio features are leveraged to estimate RIRs. It is well-established that $\mathcal{V}$ and \rts\ are significantly correlated with RIRs \cite{yu2020room}. 

Initially, we map the global acoustic features $\mathbf{q}_\mathcal{V}$ and $\mathbf{q}_\zeta$ through a Projection Head consisting of a LayerNorm and a linear layer. The module acquires room acoustic features related to $\mathcal{V}$, denoted as $\mathbf{r}_\mathcal{V} \in \mathbb{R}^{N_\mathcal{V} \times 1}$, and room acoustic features related to \rts, denoted as $ \mathbf{r}_\zeta \in \mathbb{R}^{N_\zeta \times 1} $. These acoustic features are then fused into a unified vector through concatenation. This composite vector undergoes dimension matching through a linear transformation layer, aligning it with the audio features. Given the time-invariant nature of $\mathbf{r}_\mathcal{V}$ and $\mathbf{r}_{\zeta}$, we extend this vector along the time dimension to form a high-dimensional representation of room acoustic features, $\bm{\mathbf{F}}_{\mathrm{a}} \in \mathbb{R}^{N_\mathrm{a} \times L}$, corresponding to the length $L$ of the audio features. This process follows the dimension matching strategy described in \cite{song2023conditional}, which can be mathematically expressed as:
\vspace{-3mm}
\begin{equation}
    \mathbf{F}_{\mathrm{a}}  = \text{Broadcast}(\text{Linear}_\mathrm{r}(\text{Concat}(\mathbf{r}_{\mathcal{V}}, \mathbf{r}_{\zeta}))).
    \label{Fa}
\end{equation}
\vspace{-5mm}

Existing blind RIR estimation methods often focus solely on a single category of features, which may not fully capture the diversity of real-world environments. This limitation can result in discrepancies between the estimated RIR and the actual environment. Inspired by recent research on feature fusion \cite{tao2021someone, rong2023dynstatf, zhang2023cross}, we propose a hybrid-path cross-attention feature fusion module to facilitate information interaction between $\bm{\mathbf{F}}_{\mathrm{s}}$ and $\bm{\mathbf{F}}_{\mathrm{a}}$.

As shown in Fig. \ref{fig:hpca}, the hybrid-path cross-attention module consists of two parallel paths. In the first path, we focus on exploring the guiding role of room acoustic features $\bm{\mathbf{F}}_{\mathrm{a}}$ on the audio features $\bm{\mathbf{F}}_{\mathrm{s}}$. Specifically, we use the audio features $\bm{\mathbf{F}}_{\mathrm{s}}$ as the source sequence and the room acoustic features $\bm{\mathbf{F}}_{\mathrm{a}}$ as the target sequence. Through the cross-attention layer, we generate audio attention features guided by the room acoustic features $\bm{\mathbf{F}}_{\mathrm{a}}$. The features $\bm{\mathbf{F}}_{\mathrm{s}}$ and $\bm{\mathbf{F}}_{\mathrm{a}}$ are extracted by two distinct encoders. This leads to a lack of necessary global interaction between features, limiting the depth and effectiveness of feature fusion, thus affecting the final RIR estimation performance. When only simple linear layers are employed for fusion, it is difficult for the model to achieve deeper feature interaction and integration.

To address the issues mentioned above, a pre-fusion strategy is adopted, following \cite{chen2021crossvit}. $\bm{\mathbf{F}}_{\mathrm{s}}$ and $\bm{\mathbf{F}}_{\mathrm{a}}$ are first concatenated along the feature dimension and passed through a projection function $\mathcal{P}_{\mathrm{a}}(\cdot)$, which maps the concatenated features into a unified feature space, yielding latent representations of the pre-fused features $\bm{\mathbf{X}}_{\mathrm{a}}^{\text{Pre}} \in \mathbb{R}^{N_\mathrm{a} \times L}$:
\begingroup
  \setlength{\abovedisplayskip}{6pt} 
  \setlength{\belowdisplayskip}{6pt}   
  \begin{equation}
    \bm{\mathbf{X}}_{\mathrm{a}}^{\text{Pre}}
    = \mathcal{P}_{\mathrm{a}}(\text{Concat}(\bm{\mathbf{F}}_{\mathrm{a}}, \bm{\mathbf{F}}_{\mathrm{s}})).
    \label{Xapre}
  \end{equation}
\endgroup

Here, $\text{Concat}(\cdot, \cdot)$ denotes the concatenation operation along the feature dimension. 
$\mathcal{P}_{\mathrm{a}}(\cdot)$ performs dimension alignment through a linear projection, followed by a GELU activation function: $\mathcal{P}_{\mathrm{a}}(\cdot)=\text{GELU}(\textbf{W}_\mathrm{a}(\cdot)+\mathbf{b}_\mathrm{a}),$ where $\mathbf{W}_\mathrm{a}$ and $\mathbf{b}_\mathrm{a}$ are learnable parameters.
Through pre-fusion, the room acoustic features are initially enhanced with audio information on a global scale. However, considering that simple fusion might not adequately capture the finer details of speech, we further introduce an interactive cross-attention module to achieve deeper integration of information.

As shown in Fig. \ref{fig:hpca}, the pre-fused room acoustic features $\bm{\mathbf{X}}_{\mathrm{a}}^{\text{Pre}}$ undergo a dimension permutation and are linearly mapped to form the query embeddings $\bm{\mathbf{Q}}_{\mathrm{a}} \in \mathbb{R}^{L \times d_\mathrm{k}}$ for the attention mechanism. Simultaneously, the audio features $\bm{\mathbf{F}}_{\mathrm{s}}$ are also subjected to dimension permutation and linearly mapped to generate the keys $\bm{\mathbf{K}}_{\mathrm{s}}$ and values $\bm{\mathbf{V}}_\mathrm{s}$ for the attention mechanism, specifically $\bm{\mathbf{K}}_{\mathrm{s}} \in \mathbb{R}^{L \times d_\mathrm{k}}$ and $\bm{\mathbf{V}}_{\mathrm{s}} \in \mathbb{R}^{L \times d_\mathrm{v}}$. Here, $L$ represents the sequence length, $d_\mathrm{k}$  denotes the dimension of the query and key, and $d_\mathrm{v}$  denotes the dimension of the value. The scaled dot-product cross-attention mechanism, $\mathbf{C}_{\text{Atten}}$, is defined as follows:
\vspace{-0.5em}
\begin{align}
    \bm{\mathbf{C}}_{\text{Atten}}(\bm{\mathbf{F}}_{\mathrm{s}}, \bm{\mathbf{X}}_\mathrm{a}^{\text{Pre}}) &= \text{softmax}\left(\frac{\bm{\mathbf{X}}_\mathrm{a}^{\text{Pre}} \bm{\mathbf{W}}_\mathrm{Q_a} \bm{\mathbf{W}}_\mathrm{K_s}^T \bm{\mathbf{F}}_{\mathrm{s}}^T}{\sqrt{d_\mathrm{k}}}\right) \bm{\mathbf{F}}_{\mathrm{s}} \bm{\mathbf{W}}_\mathrm{V_s} \notag \\
    &=  \text{softmax} \left( \frac{\bm{\mathbf{Q}}_\mathrm{a} \bm{\mathbf{K}}_\mathrm{s}^T}{\sqrt{d_\mathrm{k}}} \right) \bm{\mathbf{V}}_\mathrm{s}.
    \label{CAtten}
\end{align}

$\bm{\mathbf{W}}_\mathrm{Q_a}$, $\bm{\mathbf{W}}_\mathrm{K_s}$, and  $\bm{\mathbf{W}}_\mathrm{V_s}$ correspond to the projection parameters for the acoustic feature query, audio feature key, and audio feature value, respectively. The subscript $\mathrm{a}$ denotes the representation of room acoustic features and $\mathrm{s}$ denotes the representation of audio features. The function $ \text{softmax}(\cdot) $ is used for weight normalization, and  $\sqrt{d_\mathrm{k}}$ represents the normalization factor for the dimensionality of the attention.

To fully capture information across different dimensions, we employ a multi-head cross-attention mechanism. The outputs of cross-attention from different heads are concatenated and then mapped through a linear layer to obtain the multi-head cross-attention features $\bm{\mathbf{M}}_{\mathrm{C}_{\text{Atten}}}$, represented as: 
\vspace{-0.2em}
\begin{equation}
\mathbf{M}_{\mathrm{C}_{\text{Atten}}}(\mathbf{F}_\mathrm{s}, \mathbf{X}_\mathrm{a}^{\text{Pre}}) = \text{Concat}(\bm{\mathbf{C}}_{\text{Atten}_1}, \ldots, \bm{\mathbf{C}}_{\text{Atten}_H}) \mathbf{W}_\mathrm{s,a},
\end{equation}
where $H$ is the total number of heads, and  $\mathbf{W}_\mathrm{s,a}$ indicates the set of trainable parameters. $\mathbf{C}_{\text{Atten}_{1,\ldots,H}}$ are defined as the cross-attention outputs in Equation \ref{CAtten}, where the subscript indexes the head number.

Through the multi-head cross-attention, the model is able to focus on the key audio information guided by room acoustic features. To further enhance the feature representation, residual connections, feed-forward networks (FFNs), and layer normalization are introduced. Consequently, the enhanced audio attention features $\mathbf{F}_\mathrm{s}^{\text{Enh}}$ are obtained. The specific calculation process is as follows:
\begin{equation}
\mathbf{F}_\mathrm{s}^{\text{Mid}} = \text{LN}(\mathbf{F}_\mathrm{s} + \mathbf{M}_{\mathrm{C}_{\text{Atten}}}(\mathbf{F}_\mathrm{s}, \mathbf{X}_\mathrm{a}^{\text{Pre}})),
\end{equation}
\begin{equation}
\mathbf{F}_\mathrm{s}^{\text{FFN}} = \text{ReLU}(\mathbf{F}_\mathrm{s}^{\text{Mid}} \mathbf{W}_1 + \mathbf{b}_1) \mathbf{W}_2 + \mathbf{b}_2,
\label{FsFFN}
\end{equation}
\begin{equation}
\mathbf{F}_\mathrm{s}^{\text{Out}} = \text{LN}(\mathbf{F}_\mathrm{s}^{\text{FFN}} + \mathbf{F}_\mathrm{s}^{\text{Mid}}),
\end{equation}
where $\text{LN}(\cdot)$ denotes layer normalization. The matrices $ \mathbf{W}_1 $ and $ \mathbf{W}_2 $ are linear projection matrices, where $ \mathbf{W}_1 \in \mathbb{R}^{N_\mathrm{a} \times N_\mathrm{ff}}$ and $ \mathbf{W}_2 \in \mathbb{R}^{N_\mathrm{ff} \times N_\mathrm{a}}$, and $ \mathbf{b}_1, \mathbf{b}_2 $ are bias terms. Here, $ N_\mathrm{ff} = 4N_\mathrm{a} $.

Another parallel path is dedicated to extracting and enhancing room acoustic information from $\mathbf{F}_{\mathrm{a}}$. Unlike dynamic and richly detailed audio features $ \mathbf{F}_\mathrm{s} $, room acoustic features $ \mathbf{F}_{\mathrm{a}}$ are generally more static and contain only indirect information. Instead of employing complex attention mechanisms, a lightweight FFN is used to independently enhance the room acoustic features $ \mathbf{F}_{\mathrm{a}}$. The specific computation steps are as follows:
\vspace{-0.2em}
\begin{equation}
\mathbf{F}_{\mathrm{a}} ^{\text{Enh}} = \text{LN}(\mathbf{F}_{\mathrm{a}}  + \text{ReLU}(\mathbf{F}_{\mathrm{a}}  \mathbf{W}_3 + \mathbf{b}_3) \mathbf{W}_4 + \mathbf{b}_4), 
\vspace{-2mm}
\end{equation}
where $\text{LN}(\cdot)$ denotes layer normalization. The matrices $\mathbf{W}_3 \in \mathbb{R}^{N_\mathrm{a} \times N_\mathrm{ff}}$ and $\mathbf{W}_4 \in \mathbb{R}^{N_\mathrm{ff} \times N_\mathrm{a}}$ are linear projection matrices, and $\mathbf{b}_3, \mathbf{b}_4$ are bias terms.

Finally, the feature representations produced by the two parallel paths are concatenated along the feature dimension and passed through a projection function $\mathcal{P}_\mathrm{c}(\cdot)$ to obtain the fused representation $\mathbf{F}_\mathrm{c} \in \mathbb{R}^{N_\mathrm{c} \times L}$:

\begingroup
  \setlength{\abovedisplayskip}{6pt} 
  \setlength{\belowdisplayskip}{6pt}
  \vspace{-3mm}
  \begin{equation}
  \mathbf{F}_\mathrm{c} = \mathcal{P}_\mathrm{c}(\text{Concat}(\mathbf{F}_\mathrm{s}^{\text{Enh}}, \mathbf{F}_{\mathrm{a}} ^{\text{Enh}})).
  \end{equation}
\endgroup
Through this hybrid-path collaborative optimization strategy, the model is able to enhance both audio and room acoustic features, resulting in deeper feature fusion between them.

\vspace{-4mm}
\subsection{Dynamic Acoustic Tuning Decoder}
In this section, we propose a DAT decoder aimed at generating RIRs that closely match real-world environments based on room acoustic characteristics.

Conventionally, the RIR can be modeled as the combination of an early reverberation component and a late reverberation component, where the early reverberation includes both direct sound and early reflections. Typically, the late reverberation is described by an exponentially decaying filtered noise model \cite{moorer1979reverberation, steinmetz2021filtered}. Although previous studies commonly use a fixed threshold of 50 $ms$ to distinguish early and late reverberation \cite{steinmetz2021filtered}, this empirical value may not always be accurate, particularly when factors such as room dimensions and wall absorption coefficients vary significantly \cite{polack1993playing, stewart2007statistical,abel2006simple}. In scenarios that involve larger room volumes or lower wall absorption coefficients, the energy decay characteristics of the reverberant field are more likely to deviate noticeably from this fixed boundary, resulting in reduced modeling accuracy.

\begin{figure}[ht]
\centering
\includegraphics[width=8cm]{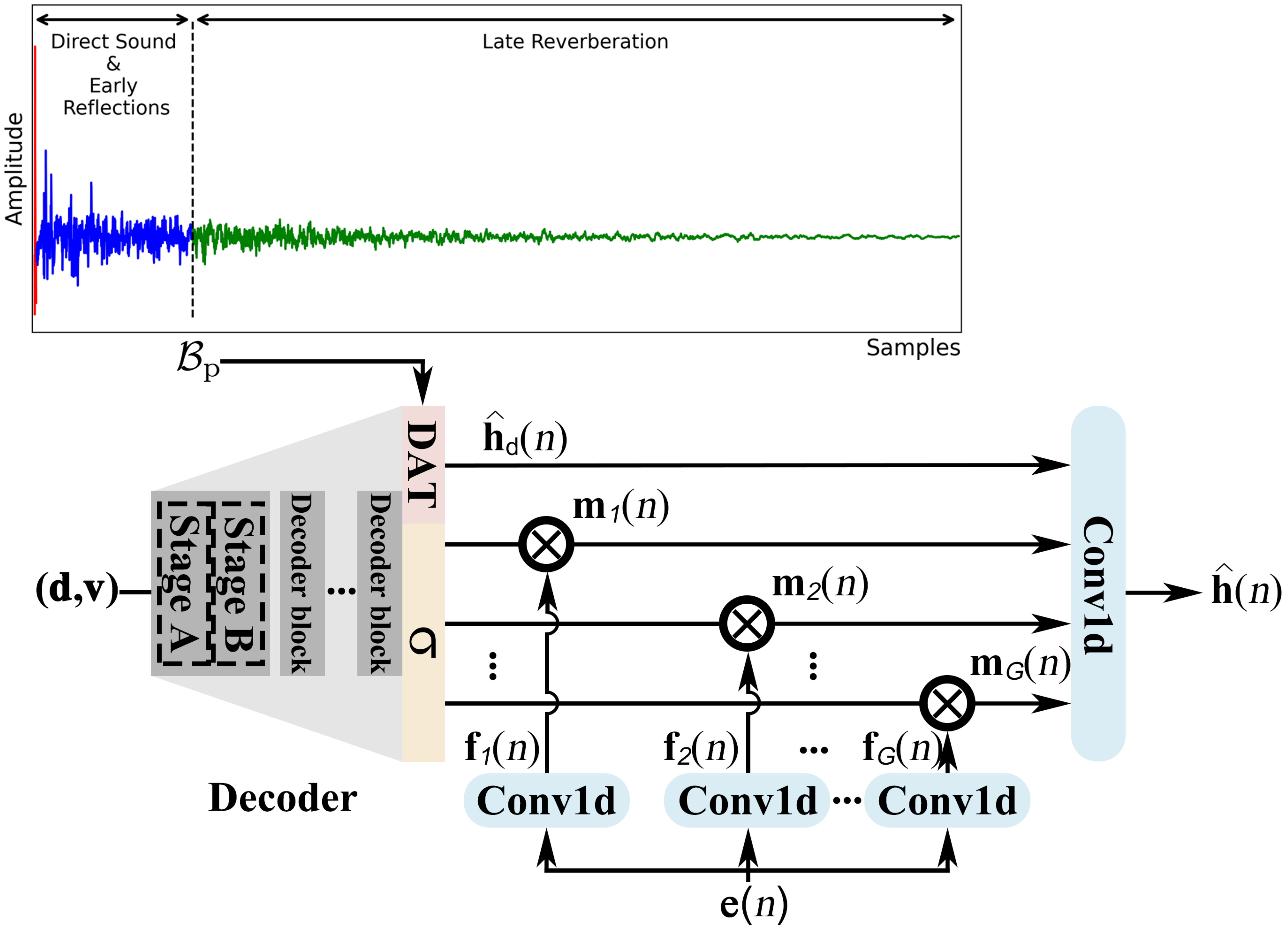}
\vspace{-3mm}
\caption{Schematic diagram of the proposed DAT decoder for dynamically modeling early and late reverberation. The decoder divides the estimated RIR $\hat{\mathbf{h}}(n)$ into an early reverberation component $\hat{\mathbf{h}}_\mathrm{d}(n)$ and a late reverberation component based on the dynamic boundary point $\mathcal{B}_\mathrm{p}$.}
\vspace{-7mm}
\label{fig:decoder}
\end{figure}

To address this issue, we propose a DAT decoder that actively adjusts the boundary point between early and late reverberation based on various room characteristics, as shown in Fig. \ref{fig:decoder}. The overall architecture of the decoder is based on the generator from GAN-TTS \cite{binkowski2019high}. Specifically, utterance information that exceeds the size of the receptive field is aggregated using pooling in the mixed features $\mathbf{F}_\mathrm{c}$, resulting in an $N_\mathrm{c}$-dimensional latent embedding $\mathbf{d}$ that encompasses key information of both audio features and  room acoustic features.  By continually injecting the latent embedding $\mathbf{d}$, the decoder is trained to generate the early reverberation component $\hat{\mathbf{h}}_\mathrm{d}(n)$ of the estimated RIR together with $G$ time-domain masks $\mathbf{m}_1(n),...,\mathbf{m}_G(n)$. These masks are then applied to a set of filtered noise signals $\mathbf{f}_1(n),...,\mathbf{f}_G(n)$, thereby producing its late reverberation.

Within the decoder, each decoder block consists of two stages and utilizes FiLM \cite{perez2018film} for conditional control. The latent embedding $\mathbf{d} \in \mathbb{R}^{1 \times N_\mathrm{c}}$ combined with the noise vector $\mathbf{v} \in \mathbb{R}^{1 \times Z}$ is continuously injected into these two stages through FiLM operations. In Stage A, features are upsampled and refined through transposed convolutions to form a coarse temporal structure. In Stage B, additional convolutions with larger dilation factors further refine the upsampled signal.


The decoder dynamically divides the estimated RIR into early and late reverberation stages. As early reverberation is not suitable for direct capture using the masking method, an additional output channel is configured for the time-domain early reverberation component $\hat{\mathbf{h}}_\mathrm{d}(n) \in \mathbb{R}^{1 \times \mathcal{B}_\mathrm{p}}$, where $\mathcal{B}_\mathrm{p}$ represents a dynamic boundary point estimated by the MASS-BRPE module, and the MASS-BRPE module is frozen during the subsequent RIR training process. For samples where $n >\mathcal{B}_\mathrm{p}$, $\hat{\mathbf{h}}_\mathrm{d}(n)$ is set to zero.

To model late reverberation, we simulate this part by taking a weighted sum of filtered noise signals. Specifically, the noise signal $\mathbf{e}(n)$ is processed through a series of trainable FIR filters to generate the filtered noise signal $\mathbf{f}_g(n)$. Each filter is implemented via 1-D convolution, and the corresponding operation can be modeled as:

\begingroup
  \setlength{\abovedisplayskip}{6pt} 
  \setlength{\belowdisplayskip}{6pt} 
\vspace{-3mm}
\begin{equation}
\mathbf{f}_g[\iota] = \sum_{\gamma=0}^{\Gamma} \lambda_{g,\gamma} \cdot \mathbf{e}[\iota-\gamma],
\vspace{-3mm}
\end{equation}
\endgroup
where $\Gamma$ denotes the order of the FIR filter, $g$ is the index of the FIR filter, ranging from $g = 1, \ldots, G$, and $G$ denotes the total number of FIR filters. $[\iota]$ represents the $\iota$-th element of a vector. $\lambda_{g,\gamma}$ is the $\gamma$-th coefficient of the $g$-th filter. 

Subsequently, the decoder generates several time domain masks $\mathbf{m}_g \in \mathbb{R}^{1 \times U}$ to shape the filtered noise signals $\mathbf{f}_g \in \mathbb{R}^{1 \times U}$. Each subband component is defined as:
\begin{equation}
\hat{\mathbf{h}}_{u, g} = \mathbf{f}_g(n) \odot \sigma(\mathbf{m}_g(n)),
\end{equation}
where $\sigma(\cdot)$ denotes the sigmoid function, and $\odot$ represents element-wise multiplication. Finally, the estimated early reverberation component $\hat{\mathbf{h}}_\mathrm{d}(n)$ is mixed with each sub-band component of $G$ late reverberation parts $\hat{h}_{u, g}$, and processed through a $1 \times 1$ convolution to produce the estimated monaural RIR $\hat{\mathbf{h}}(n)$.

\vspace{-4mm}
\subsection{Loss Function}

The structural characteristics of the RIR vary between its components. Specifically, the early part of the RIR often exhibits sparse impulsive features, while the latter part resembles noise. To estimate the monaural RIR, denoted as $\hat{\mathbf{h}}(n)$, we utilize the multi-resolution short-time Fourier transform (STFT) loss. Similar to previous work \cite{arik2018fast, yamamoto2020parallel, steinmetz2021filtered}, we define the loss function as follows:
\begingroup
  \setlength{\abovedisplayskip}{6pt} 
  \setlength{\belowdisplayskip}{6pt} 
    \begin{equation}
    \mathcal{L}_\mathrm{STFT}(\mathbf{h}, \hat{\mathbf{h}}) = \frac{1}{\Xi}\sum_{\xi=1}^{\Xi}(\mathcal{L}_{\mathrm{SC}_{(\xi)}}(\mathbf{h}, \hat{\mathbf{h}}) + \mathcal{L}_{\mathrm{MAG}_{(\xi)}}(\mathbf{h}, \hat{\mathbf{h}})), 
    \vspace{-3mm}
    \end{equation}
\endgroup
where $\mathbf{h}$ and $\hat{\mathbf{h}}$ denote the ground-truth and predicted RIRs, respectively. $\Xi$ denotes the number of STFT resolutions. $\mathcal{L}_{\mathrm{SC}_{(\xi)}}$ and $\mathcal{L}_{\mathrm{MAG}_{(\xi)}}$ represent the spectral convergence loss and the log STFT magnitude loss at the $\xi$-th STFT resolution, defined as follows:

\begin{equation}
\mathcal{L}_{\mathrm{SC}_{(\xi)}}(\mathbf{h}, \mathbf{\hat{h}}) = \frac{\| |\text{STFT}_{\xi}(\mathbf{h}) - \text{STFT}_{\xi}(\mathbf{\hat{h}})| \|_\mathrm{F}}{\| |\text{STFT}_{\xi}(\mathbf{h})| \|_\mathrm{F}},
\end{equation}
\begin{equation}
\mathcal{L}_{\mathrm{MAG}_{(\xi)}}(\mathbf{h}, \mathbf{\hat{h}}) = \frac{1}{\Psi} \|\ln(|\text{STFT}_{\xi}(\mathbf{h})|) - \ln(|\text{STFT}_{\xi}(\mathbf{\hat{h}})|) \|_1,
\end{equation}
where $\|\cdot\|_\mathrm{F}$ and $\|\cdot\|_1$ denote the Frobenius and $L_1$ norms, respectively. $\Psi$ represents the number of STFT frames.

\vspace{-2mm}
\section{Experimental Procedure}\label{Experimental Procedure}

The evaluation of our work includes seven publicly available real-world RIR datasets, supplemented by virtual room simulation data, to construct a highly realistic and diverse RIR dataset. Each RIR is annotated with its ground-truth acoustic parameters. To ensure an accurate assessment of model generalization, a subset containing only unseen real-world room RIRs is specifically selected for testing the blind RIR estimation task. Both training and testing phases standardize the sampling rate and the target RIR length, with logarithmic mapping effectively mitigating scale discrepancies between room parameters.

\begin{figure}[ht]
\centering
\includegraphics[width=3in]{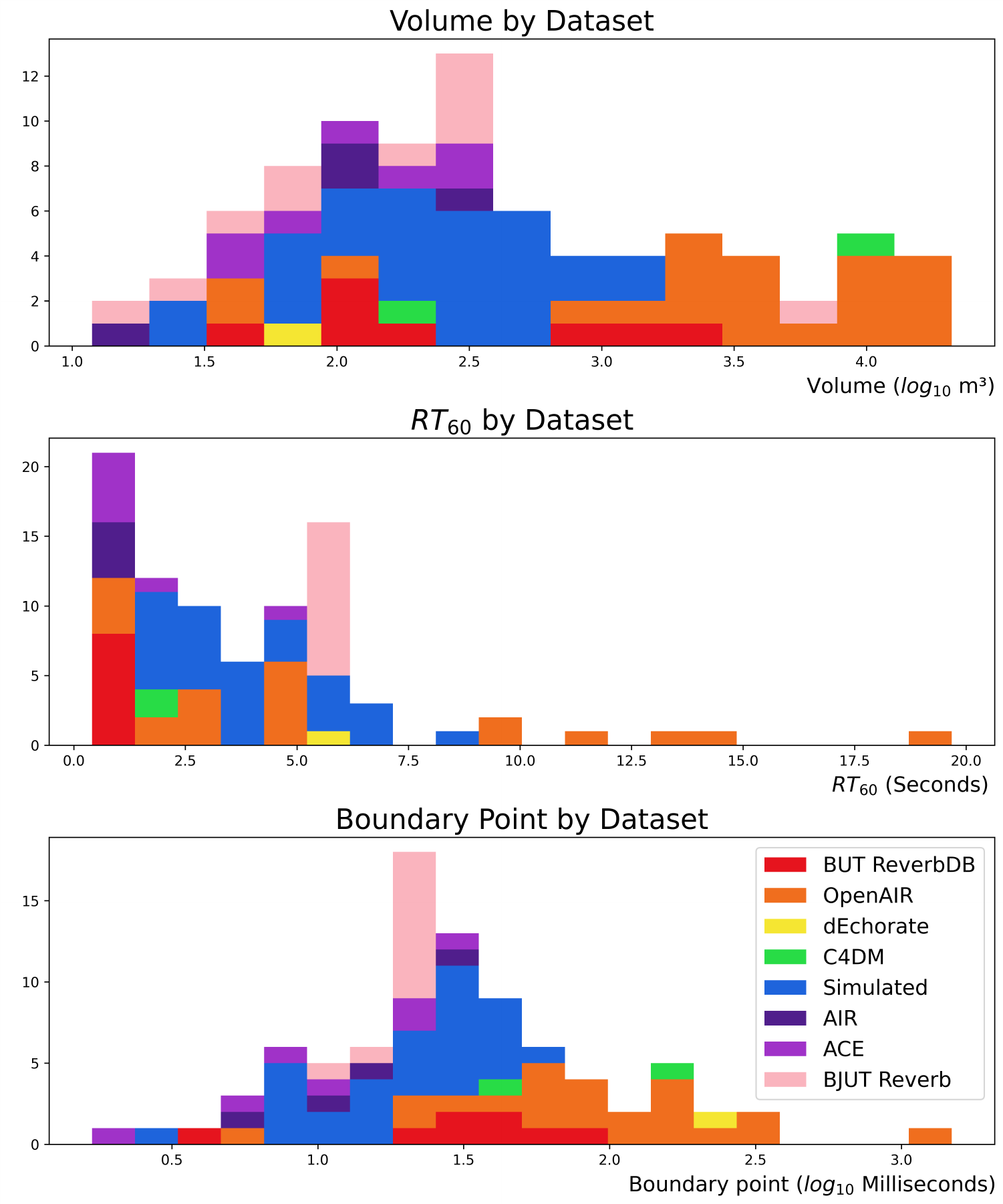}
\vspace{-4mm}
\caption{Parameter Distributions of the Constructed RIR Dataset, showing the labeled $\mathcal{V}$ (in $m^3$, on a logarithmic scale), \rts\ (in seconds), and $\mathcal{B}_\mathrm{p}$ (in milliseconds, on a logarithmic scale). }
\vspace{-3mm}
\label{fig:histogram}
\end{figure}

To precisely assess the performance of RIR estimation, a multi-resolution STFT loss function covering various typical temporal scales and frequency resolutions is employed. Additionally, multiple statistical evaluation metrics from both logarithmic and linear scales are introduced, allowing comprehensive assessment of different blind room parameter estimation (BRPE) modules' performances and facilitating analysis of their implications for subsequent RIR estimation tasks. Key acoustic parameters, including \rts\ and DRR, are further extracted to systematically investigate differences between model estimations and measured RIRs, thereby providing deeper insights into the reliability and effectiveness of model estimations in practical acoustic environments.

\vspace{-4mm}
\subsection{Dataset}

Estimating RIRs from reverberant speech using neural network methods is a highly challenging task that typically requires extensive and diverse data support. To more accurately estimate RIRs that match real-world environments, maximizing the diversity of real-world RIR samples is crucial. This study comprehensively considers seven publicly available real-world RIR datasets, encompassing 55 real rooms, aiming to cover the acoustical characteristics of real-world rooms as comprehensively as possible. These data primarily come from spaces with varying geometric structures, including auditoriums, classrooms, seminar rooms, cathedrals, and other such venues. 

Specifically, the datasets used in this study include the Brno University of Technology Reverb (BUT ReverbDB) dataset \cite{szoke2019building}, the OpenAIR dataset \cite{murphy2010openair}, the dEchorate dataset \cite{carlo2021dechorate}, the C4DM dataset \cite{stewart2010database}, the Aachen Impulse Response (AIR) dataset \cite{jeub2009binaural}, the Acoustic Characterization of Environments Challenge (ACE) dataset \cite{eaton2016estimation}, and the Beijing University of Technology Reverb (BJUT Reverb) dataset \cite{wang2024attention}. 

Additionally, to further diversify the geometric characteristics of rooms in the dataset, virtual room simulation technology is employed. In these simulations, the absorption coefficients of each room surface (including the walls, ceiling, and floor) are chosen based on the sound absorption properties of the materials in the room, with the absorption coefficients being independent of frequency. Additionally, the microphones and sound sources used are omnidirectional. Using the pyroomacoustics \cite{scheibler2018pyroomacoustics} software package based on the image source model, an additional 30 simulated room RIRs with varying geometric structures are generated.  This is done to augment areas of room volume that are less commonly represented in the actual data, thus making the overall room volume distribution of the dataset more closely approximate a normal distribution.

As shown in Fig. \ref{fig:histogram}, the constructed RIR dataset includes annotations for $\mathcal{V}$, \rts, and $\mathcal{B}_\mathrm{p}$. Among them, \rts\ values show significant variations, ranging from less than half a second to over ten seconds. These values are calculated according to the Schroeder method \cite{schroeder1965new}. As for $\mathcal{B}_\mathrm{p}$, we determine it by analyzing the echo density function $\mathrm{NED}(n)$ of the RIR \cite{abel2006simple}. It is defined as follows:
\vspace{-0.5em}
\begin{equation}
\mathrm{NED}(n) = \frac{1}{\text{erfc}\left({1}/{\sqrt{2}}\right)} \sum_{l=n-\delta}^{n+\delta} \omega(l)  \mathbf{1}\cdot\{|\textbf{h}(l)| > \varsigma\},
\vspace{-2mm}
\end{equation}
where $\frac{1}{\text{erfc}(1/\sqrt{2})}$ represents the fraction of samples outside the mean standard deviation that satisfy a Gaussian distribution. $\mathbf{1}\cdot$ is an indicator function that takes the value of 1 when $\mathbf{h}(l) > \varsigma$, and 0 otherwise. $\omega(l)$ is a weighting function, and $\varsigma$ is the standard deviation of the described RIR $\textbf{h}(l)$ in the current window. As the reverberation time and the degree of scattering increase, the NED gradually increases. The value of NED starts from 0 and approaches 1. The point where NED first remains constant is defined as the boundary between early reflections and late reverberation.

The speech samples are taken from the ACE dataset \cite{eaton2016estimation}, which contains recordings of male and female speakers' anechoic speech. Using an RIR dataset labeled with room parameters, we convolve clean speech signals recorded in an anechoic chamber to generate reverberant speech sequences with corresponding RIR characteristics. To ensure a fair comparison with existing methods, noise has not yet been considered in this work. The data are divided into training, validation, and testing sets in a 6-2-2 ratio based on the room. It is important to note that the test set only includes RIRs from real-world rooms and ensures that these RIRs are unseen during the model training process.

\vspace{-4mm}
\subsection{Training Details}

All models process input and output audio at a sampling rate of 16 kHz. The length of the target RIR is uniformly set to 1 second, corresponding to 16,000 samples. During training, we generate reverberant speech by sampling anechoic speech and RIRs from the training set, producing inputs of 80,000 samples (5 seconds at 16 kHz). Appropriate zero-padding or cropping is applied to ensure consistent data lengths.
In the hybrid-path cross-attention feature fusion, we configure a network that contains one transformer layer, with the number of attention heads $H$ set to 8. The feature dimensions for room acoustics, $N_\mathcal{V}$ and $N_\zeta$, which are related to $\mathcal{V}$ and \rts\ respectively, are initially set to 64. An ablation study is carried out with different values (see Section \ref{sec:Ablation Study}).

During the training process, each model undergoes 1,000 epochs with a batch size of 16. We use the AdamW optimizer with an initial learning rate of $5.5\mathrm{e}{-5}$, which decays 80\% of its value every 80 epochs. To stabilize the training process, we employ gradient clipping to restrict gradient norms to a maximum value of 5 and utilize automatic mixed-precision techniques to improve training efficiency.

To effectively address the significant scale differences among room parameters, we propose a logarithmic mapping method. Specifically, we apply logarithmic scaling to $\mathcal{V}$ and $\mathcal{B}_\mathrm{p}$. Additionally, \rts\ values are mapped onto volume parameters and subsequently log-scaled. It should be noted that this data processing method is reversible, allowing recovery of processed parameters to their original standard units. Compared to conventional normalization techniques, our logarithmic mapping method effectively avoids errors caused by scale differences during feature fusion, fundamentally resolving inconsistencies in units and orders of magnitude.

For the loss function, we utilize a multi-resolution STFT loss with a resolution setting of $\Xi$=4. The selected frame lengths are $32\ (2\ ms)$, $256\ (16\ ms)$, $1024\ (64\ ms)$, and $4096\ (256\ ms)$, effectively covering multiple typical time scales in RIRs. The step size for each resolution is set to half the respective frame length.

To ensure fairness in evaluating model performance, all experiments are conducted on a computing platform equipped with an Intel Core i9 processor and an NVIDIA GeForce RTX 4090 GPU.

\vspace{-4mm}
\subsection{Evaluation Metrices}

To comprehensively and accurately evaluate the discrepancies between estimated and ground-truth room parameters in the BRPE system, multiple statistical metrics on a logarithmic scale (log-10) are employed, including mean squared error (MSE), mean absolute error (MAE), Pearson correlation coefficient ($\rho$), and mean mult (MM).

Among them, the MM metric measures the exponential of the mean absolute logarithm of the ratio between the estimated and target room parameters. For instance, for the estimated volume parameter, given the ground truth volume $\mathcal{V}_i$ and the estimated volume $\hat{\mathcal{V}}_i$, it can be defined as:

\vspace{-3mm}
\begin{equation}
\text{MM} = \exp \left( \frac{1}{I} \sum_{i=1}^{I} \left| \ln \left( \frac{\hat{\mathcal{V}}_i}{\mathcal{V}_i} \right) \right| \right),
\vspace{-2mm}
\end{equation}
where $i$ represents the sample index and $I$ represents the total number of samples. This metric summarizes mean error in terms of the ratio between estimated and target parameters, using a method that is less sensitive to overall scale, thus providing more balanced weighting across different scales.

Given the wide numerical range typically encountered in real-world room parameters, the employment of a logarithmic scale effectively mitigates disproportionate influences caused by large numerical errors, ensuring a balanced evaluation across different orders of magnitude. Additionally, the median and MAE of estimation errors are reported on a linear scale to enhance interpretability and to clearly reflect model reliability in real-world scenarios.

For RIR estimation, we further analyze the model's adaptability to realistic acoustic conditions by evaluating the similarity and matching degree between the predicted RIR and the actual RIR. Specifically, we introduce $\mathcal{L}_\mathrm{STFT}$, which can simultaneously measure subtle discrepancies across multiple temporal window lengths and frequency resolutions. Such a multi-scale evaluation approach sensitively captures estimation errors across various frequency bands and time scales, effectively highlighting the model’s capability to reconstruct realistic time-frequency acoustic characteristics. In addition, we incorporate the MAE for comparative analysis to further validate the accuracy of the estimated RIR.

In addition, we conduct detailed assessments that specifically focus on key acoustic parameters within RIRs, namely \rts\ and DRR. For these critical parameters, we employ $\rho$ to evaluate the linear correlation between the estimated and true values, use the MSE and root mean squared error (RMSE) to quantify the magnitude and variability of estimation errors, and calculate the overall bias to identify systematic tendencies of over- or under-estimation. Moreover, given the large variation in \rts\ across experimental conditions, the \rts\ estimation error is defined as the signed relative percentage deviation between the ground-truth RIR $\mathbf{h}$ and the predicted RIR $\hat{\mathbf{h}}$:
\begin{equation}
{RT}_{60}\text{-Error} \;=\; 
\frac{{RT}_{60}(\hat{\mathbf{h}}) - {RT}_{60}(\mathbf{h})}{{RT}_{60}(\mathbf{h})} \times 100\%,
\label{eq:rt60-err}
\end{equation}
where ${RT}_{60}(\cdot)$ denotes the reverberation time computed from an RIR using the Schroeder backward integration method \cite{schroeder1965new}. Using these combined metrics ensures a more complete and precise evaluation of the estimated RIR performance in capturing essential acoustic attributes such as \rts\ and DRR, further validating the efficacy and reliability of the model in accurately representing realistic acoustic environments.

\section{EXPERIMENTAL RESULTS}
\label{EXPERIMENTAL RESULTS}

In this section, we compared the proposed DARAS framework, combined with the MASS-BRPE module, with other competitive baselines for the BRPE and blind RIR estimation tasks. We also provided qualitative analyses and visualization results. In addition, a MUSHRA-based subjective listening study was conducted to evaluate the consistency and fidelity of the generated RIRs against the ground truth.

\vspace{-4mm}
\subsection{Comparison with BRPE}

To verify the effectiveness of the proposed MASS-BRPE module, we selected mainstream BRPE baselines for accuracy comparison. All baselines were implemented with the same ``+\textit{Phase}" strategy \cite{ick2023blind,wang2025ss}.

(1) CNN-based module: This method is based on a feed-forward 2-D convolutional neural network that effectively extracts 2-D features from audio data. 

(2) SS-BRPE module: This approach combines a pure attention mechanism with a self-supervised learning framework, primarily using an audio spectrogram Transformer \cite{gong2021ast}. 

We evaluated the performance of CNN-based, SS-BRPE, and the proposed MASS-BRPE modules by comparing the estimated room parameters ($\mathcal{V}$, \rts,  $\mathcal{B}_\mathrm{p}$) with the ground-truth values. Detailed experimental results are shown in Table \ref{tab:EvaBRPE}, with the best results highlighted in bold.
The data indicate that the MASS-BRPE module significantly outperforms other baselines in estimating room parameters. 

On a linear scale, the MASS-BRPE module consistently outperforms all baselines, delivering predictions that more closely mirror real-world acoustic conditions.



Subsequently,we compared the inference speed and GPU memory usage of the baselines and the proposed MASS-BRPE module to comprehensively evaluate the deployment performance.

As shown in Table \ref{tab:EvaBRPE}, our proposed MASS-BRPE module achieves superior accuracy and consistency in estimating room parameters, while also demonstrating significant advantages in computational efficiency. Specifically, although the SS-BRPE module exhibits strong estimation performance, its purely attention-based architecture results in high GPU memory usage (5.20 GB) and a significantly longer average inference time (0.8005 $s$). In contrast, the MASS-BRPE module has an inference time of only 0.0768 $s$, approximately 90.41\% faster than the SS-BRPE module, significantly enhancing its processing speed in the blind RIR estimation task. Furthermore, the GPU memory usage of the MASS-BRPE module is only 2.28 GB, substantially lower than that of the SS-BRPE module.

\begin{table*}[ht!]
  \centering
  \caption{Comparative Performance and Computational Resource Consumption of MASS-BRPE and Baseline Systems.}
  \vspace{-3mm}
  \label{tab:EvaBRPE}
  \begin{tabular}{c|c|cccc|cc|c|c}
    \hline
    & &\multicolumn{6}{c|}{\textbf{Evaluation Metrics}} & \textbf{GPU Memory} & \textbf{Average Inference} \\
    \cline{3-8}
    \textbf{BRPE System} & \textbf{Estimation Type} & \multicolumn{4}{c|}{\textbf{Logarithmic Scale}} & \multicolumn{2}{c|}{\textbf{Linear Scale}} & \textbf{Usage} & \textbf{Time } \\
    \cline{3-8}
    & & \textbf{MSE} $\downarrow$ & \textbf{MAE} $\downarrow$ & \textbf{$\rho$} $\uparrow$ & MM $\downarrow$ & \textbf{Median} $\downarrow$ & \textbf{MAE} $\downarrow$ & \textbf{(GB)} & ($s$) \\
    \hline
    \multirow{3}{*}{CNN \cite{ick2023blind}} & $\mathcal{V}$ & 0.3863 & 0.4837 & 0.6984 & 3.0532 & 465.22 $m^3$ & 2239.12 $m^3$ & \multirow{3}{*}{1.81} & \multirow{3}{*}{0.0465} \\
    & \rts & 0.1473 & 0.2966 & 0.8817 & 1.9952 & 0.25 $s$ & 1.60 $s$ & & \\
    & $\mathcal{B}_\mathrm{p}$ & 0.0983 & 0.2508 & 0.8113 & 1.6403 & 7 $ms$ & 11.49 $ms$ & & \\
    \hline
    \multirow{3}{*}{SS-BRPE \cite{wang2025ss}} & $\mathcal{V}$ & 0.2003 & 0.2887 & 0.8937 & 1.9599 & 234.47 $m^3$ & 1532.32 $m^3$ & \multirow{3}{*}{5.20} & \multirow{3}{*}{0.8005} \\
    & \rts & 0.0479 & 0.1470 & 0.9633 & 1.4029 & 0.09 $s$ & 0.49 $s$ & & \\
    & $\mathcal{B}_\mathrm{p}$ & 0.0376 & 0.1839 & 0.9231 & 1.5412 & 5 $ms$ & 9.03 $ms$ & & \\
    \hline
    \multirow{3}{*}{MASS-BRPE} & $\mathcal{V}$ & \textbf{0.1894} & \textbf{0.2704} & \textbf{0.8969} & \textbf{1.9496} & \textbf{226.04 $m^3$} & \textbf{1440.93 $m^3$} & \multirow{3}{*}{2.28} & \multirow{3}{*}{0.0768} \\
    & \textbf{\rts} & \textbf{0.0388} & \textbf{0.1406} & \textbf{0.9727} & \textbf{1.4021} & \textbf{0.09 $s$} & \textbf{0.40 $s$} & & \\
    & $\mathcal{B}_\mathrm{p}$ & \textbf{0.0242} & \textbf{0.1214} & \textbf{0.9513} & \textbf{1.3225} & \textbf{3 $ms$} & \textbf{6.34 $ms$} & & \\
    \hline
  \end{tabular}
  \vspace{-4mm}
\end{table*}

\vspace{-5mm}
\subsection{Comparison with Blind RIR Estimation Baseline Models}

To evaluate the effectiveness of the proposed DARAS blind RIR estimation model, we selected several representative blind RIR estimation baseline models from current research for comparative analysis and thoroughly discussed their advantages and limitations.

(1) Wave-U-Net \cite{stoller2018wave}: To tailor the original time-domain source-separation architecture Wave-U-Net to the blind RIR estimation task, we introduce several targeted modifications. Following the adjustments proposed in \cite{steinmetz2021filtered}, we largely retain the 12-layer symmetric convolutional encoder–decoder structure, while replacing the linear up-sampling blocks in the decoder with transposed convolutions to enhance the reconstruction of fine-grained temporal details.

(2) FiNS \cite{steinmetz2021filtered}: This method adopts a filtered noise shaping approach for direct RIR estimation and is primarily composed of a time-domain convolutional encoder and a specially designed decoder. The encoder gradually downsamples the input signal through multiple layers of 1-D convolution, effectively capturing temporal features at various scales. The decoder reconstructs the RIR precisely by utilizing multiple exponentially decaying filtered noise signals combined with direct sound and early reflection components.

(3) BERP \cite{wang2024berp}: This method employs a hybrid approach that stochastically models the RIR with only a few control parameters. Specifically, it employs the sparse stochastic impulse response (SSIR) model, which splits the RIR energy envelope into an onset transition and an exponential decay. The SSIR model uses a uniform Poisson process to capture the sparsity of early reflection arrivals, with sparsity approximately proportional to the room volume, whereas the exponential decay is modeled as a centered Gaussian process. These two segments are governed by rise and decay time constants that control their respective envelopes.

(4) Dbre \cite{yapar2024demucs}: A blind RIR estimation model with a convolutional–recurrent encoder–bottleneck–decoder architecture, inspired by \cite{lu2025rir, yapar2024demucs}, is introduced. The model first extracts local features from reverberant speech using convolution and downsampling. The bottleneck utilizes a two-layer long short-term memory (LSTM) to capture long-term dependencies. The decoder then reconstructs high-resolution temporal signals using upsampling, further improving estimation accuracy via a multi-resolution STFT loss.

Furthermore, to analyze the impact of the BRPE accuracy on the performance of the DARAS framework, we performed comparative experiments with multiple BRPE modules, including CNN \cite{ick2023blind}, SS-BRPE \cite{wang2025ss}, MASS-BRPE and Ground Truth (actual room parameters). Notably, all aforementioned methods require only reverberant speech during inference, making them suitable for non-intrusive applications, while Ground Truth serves as an ideal reference for evaluating the potential and performance of deep learning models.

Table \ref{tab:EvaRIR} presents detailed results that compare various baseline models and the DARAS framework in multiple objective metrics. The $\mathcal{L}_\mathrm{STFT}$ values are means with 95\% confidence intervals.
\begin{table*}[ht]
\centering
\caption{Performance Comparison of DARAS and Baseline Models on Blind RIR Estimation Metrics}
\vspace{-3mm}
\label{tab:EvaRIR}
\begin{tabular}{@{}l@{\hskip -0.031in}c@{\hskip 0.075in}c@{\hskip 0.075in}c@{\hskip 0.075in}c@{\hskip 0.075in}c@{\hskip 0.075in}c@{\hskip 0.075in}c@{\hskip 0.075in}c@{\hskip 0.075in}c@{\hskip 0.075in}c@{\hskip 0.055in}c@{\hskip 0.005in}c@{\hskip 0.005in}c@{\hskip 0.005in}c@{\hskip 0.005in}c@{}}
\toprule
\multirow{3}{*}{Model} & \multirow{3}{*}{BRPE} & \multirow{2}{*}{$\mathcal{L}_\mathrm{STFT}$} & \multirow{2}{*}{ $\mathcal{L}_\mathrm{MAG}$} & \multirow{2}{*}{$\mathcal{L}_\mathrm{SC}$} & MAE &  & \rts & & & & & DRR \\
\cline{7-10} \cline{12-14}
 & & & & & ($\times10^{-2}$) & $\rho$ & MSE ($s$) & Bias & Error ($\%$) & & $\rho$ & RMSE ($dB$) & Bias \\ 
  & & $\downarrow$ & $\downarrow$ & $\downarrow$ & $\downarrow$ & $\uparrow$ & $\downarrow$ & $|\downarrow|$ & $|\downarrow|$ & & $\uparrow$ & $\downarrow$ & $|\downarrow|$ \\ \midrule
Wave-U-Net \cite{stoller2018wave} & / & 2.1553 {\scriptsize $\pm$0.0284} & 1.2814 & 0.8739 & 1.5454 & 0.3804 & 0.3772 & 0.2447 & 21.35 & & 0.6831 & 7.9674 & -6.0391 \\
FiNS \cite{steinmetz2021filtered} & / & 1.8342 {\scriptsize $\pm$0.0264} & 1.0173 & 0.8169 & 1.5342 & 0.7166 & 0.3301 & 0.2558 & 14.13 & & 0.7241 & 7.3685 & -4.9556 \\
BERP \cite{wang2024berp} & / & 3.4911 {\scriptsize $\pm$0.0296} & 0.9472 & 2.5439 & 1.7312 & \textbf{0.9214} & \textbf{0.0723} & \textbf{0.1472} & \textbf{-6.73} & & 0.7532 & 6.3426 & -4.2642 \\
Dbre \cite{yapar2024demucs} & / & 1.6643 {\scriptsize $\pm$0.0213} & 0.8642 & 0.8001 & 1.3426 & 0.7821 & 0.1892 & 0.2142 & 9.65 & & 0.7641 & 5.4618 & -4.3288 \\
\hline
\multirow{5}{*}{DARAS} & CNN \cite{ick2023blind} & 1.7851 {\scriptsize $\pm$0.0221} & 0.9028 & 0.8823 & 1.4987 & 0.7989 & 0.2438 & 0.2291 & 10.46 & & 0.7569 & 6.0723 & -4.3149 \\
 & SS-BRPE \cite{wang2025ss} & \textbf{1.3972 {\scriptsize $\pm$0.0174}} &\textbf{0.7021} & \textbf{0.6951} & \textbf{1.2746} & \textbf{0.9012} & \textbf{0.1031} & \textbf{-0.1923} & \textbf{-6.42}  & & \textbf{0.7712} & \textbf{4.5806}& \textbf{-3.3021} \\
 & MASS-BRPE & \textbf{1.3695 {\scriptsize $\pm$0.0151}} & \textbf{0.6951} & \textbf{0.6744} & \textbf{1.1252} & \textbf{0.9321} & \textbf{0.0988} & \textbf{-0.1732 }& \textbf{-5.86} & & \textbf{0.7785} & \textbf{4.4105} & \textbf{-3.2607} \\
 \hhline{~=============}
 
 &Ground Truth & \textbf{1.1267 {\scriptsize $\pm$0.0149}} & \textbf{0.5792} & \textbf{0.5475} & \textbf{1.0028} & \textbf{0.9686} & \textbf{0.0334} & \textbf{-0.1167} & \textbf{-4.24}  & & \textbf{0.8065} & \textbf{3.1372} & \textbf{-2.1173} \\
\bottomrule
\end{tabular}
\vspace{-4mm}
\end{table*}
The results indicate that Wave-U-Net is capable of reproducing time-frequency details to some extent, predicting \rts\, and describing the DRR. However, the model faces challenges in performing blind RIR estimation in acoustically complex real-world scenarios. Although FiNS is structurally simpler and effectively captures certain temporal features, it is slightly inferior in metrics such as $\mathcal{L}_\mathrm{MAG}$, $\mathcal{L}_\mathrm{SC}$, and MAE, which directly reflect RIR estimation accuracy, indicating some distortion compared to real conditions. Furthermore, FiNS exhibits substantial estimation errors for \rts\ and DRR, with particularly pronounced errors observed in DRR, highlighting its shortcomings in complex real-world acoustic environments. Although BERP can estimate \rts\ with reasonable accuracy by predicting control parameters, certain deviations can still be observed in metrics such as $\mathcal{L}_\mathrm{STFT}$, $\mathcal{L}_\mathrm{SC}$, thereby indicating a limited ability to capture all the complexities of RIRs. The underlying reason is that BERP essentially remains a non-end-to-end paradigm based on parametric RIR synthesis and has yet to directly establish a mapping between the ground-truth and predicted RIRs.

In contrast, the Dbre model demonstrates notable improvements, particularly in metrics such as $\mathcal{L}_\mathrm{STFT}$ and MAE, showcasing stronger time-frequency reconstruction capabilities.
The Dbre model’s combined CNN and LSTM U-Net structure effectively captures local and global features, achieving $\rho$ values of 0.7821 and 0.7641 for \rts\ and DRR, respectively, indicating its estimated RIRs are closer to real conditions but still exhibit a gap.

Furthermore, the proposed DARAS framework is combined separately with three different BRPE modules: CNN-based, SS-BRPE, and MASS-BRPE. Experimental results clearly demonstrate that the DARAS framework, leveraging effective fusion of audio and room acoustic features, significantly enhances RIR estimation accuracy. We conduct pairwise $t$-tests comparing the proposed DARAS framework with each baseline model. All differences are found to be statistically significant at the $\alpha = 0.05$ level. Additionally, as the precision of room acoustic parameters extracted by the BRPE module improves, the DARAS framework exhibits an evident upward trend across various objective metrics. Although  the CNN-based DARAS framework exhibits limited parameter estimation capabilities, its generated RIR metrics already surpass FiNS and show promising results closer to real environments in terms of \rts\ and DRR. 


\begin{table*}[ht]
\centering
\caption{Ablation Study of the DARAS Model with MASS-BRPE System}
\vspace{-3mm}
\label{tab:AS}
\begin{tabular}{c c c|c c c c|c c c c}
\hline
\multirow{3}{*}{\textbf{Fusion Method}} 
 & \multirow{3}{*}{\textbf{Type}} 
 & \multirow{3}{*}{\textbf{Deep Audio Encoder}} 
 & \multicolumn{4}{c|}{\textbf{Room Acoustic Features \& Parameter}} 
 & \multirow{3}{*}{$\mathcal{L}_\mathrm{STFT}$ $\downarrow$} 
 & \multirow{3}{*}{$\mathcal{L}_\mathrm{MAG}$ $\downarrow$} 
 & \multirow{3}{*}{$\mathcal{L}_\mathrm{SC}$ $\downarrow$}
 & \multirow{3}{*}{%
  \shortstack{
    \textbf{MAE} \\
    $(\times10^{-2})$
     }
   }
\\
\cline{4-7}
 &  &  & \multirow{2}{*}{$dim( N_\mathcal{V})$} 
 & \multirow{2}{*}{$dim( N_\mathcal{\zeta})$)} 
 & \multicolumn{2}{c|}{\textbf{Boundary Point}} 
 &  &  &  & 
\\
\cline{6-7}
 &  &  &  &  & \textbf{50 $ms$} & $\mathcal{B}_\mathrm{p}$ 
 &  &  &  & 
\\
\hline
\multirow{7}{*}{\shortstack{Naive \\  \\  Feature-level \\ \\ Fusion}}
 & A & /   & 16  & 16  & \checkmark & /          & 1.7788 & 0.8965 & 0.8823 & 1.4873 \\
 & B & /   & 32  & 32  & \checkmark & /          & 1.7521 & 0.8932 & 0.8589 & 1.4926 \\
 & C & /   & 64  & 64  & \checkmark & /          & 1.7443 & 0.8892 & 0.8551 & 1.4375 \\
 & D & /   & 128 & 128 & \checkmark & /          & 1.7632 & 0.9042 & 0.8590 & 1.5021 \\
 & E & /   & 64  & 64  & /          & \checkmark & 1.6597 & 0.8431 & 0.8166 & 1.3564 \\
 & F & \checkmark & 64 & 64  & \checkmark & /       & 1.6746 & 0.8573 & 0.8173 & 1.3832 \\
 & G & \checkmark & 64 & 64  & /          & \checkmark & 1.6037 & 0.8254 & 0.7783 & 1.3025 \\
\hline
\multirow{7}{*}{\shortstack{Hybrid-Path \\ \\ Cross-Attention \\ \\ Feature Fusion}}
 & A & /   & 16  & 16  & \checkmark & /          & 1.6032 & 0.8097 & 0.7935 & 1.3025 \\
 & B & /   & 32  & 32  & \checkmark & /          & 1.5942 & 0.8126 & 0.7816 & 1.2932 \\
 & C & /   & 64  & 64  & \checkmark & /          & 1.5721 & 0.8112 & 0.7609 & 1.2357 \\
 & D & /   & 128 & 128 & \checkmark & /          & 1.5841 & 0.7998 & 0.7843 & 1.2874 \\
 & E & /   & 64  & 64  & /          & \checkmark & 1.4232 & 0.7297 & 0.6935 & 1.2095 \\
 & F & \checkmark & 64 & 64  & \checkmark & /       & 1.4527 & 0.7289 & 0.7238 & 1.1386 \\
 & G & \checkmark & 64 & 64  & /          & \checkmark & 1.3695 & 0.6951 & 0.6744 & 1.1252 \\
\hline
\end{tabular}
\vspace{-5mm}
\end{table*}

Notably, the DARAS framework combined with the MASS-BRPE module achieves the best performance due to its superior parameter estimation accuracy and powerful feature fusion capabilities, significantly integrating audio and room acoustic information. It demonstrates the highest accuracy in $\mathcal{L}_\mathrm{STFT}$, \rts, and DRR estimation, with $\rho$ values reaching 0.9321 and 0.7785 for \rts\ and DRR, respectively, confirming the DARAS framework's capability to realistically and effectively model actual room acoustic characteristics.

It is not surprising that Ground Truth serves as the performance ceiling for non-intrusive methods since obtaining actual measurements is infeasible in realistic non-intrusive scenarios. However, Ground Truth metrics indicate the existing gap between current non-intrusive estimation methods and actual measurements. These findings suggest that narrowing this gap can significantly enhance accuracy and robustness in modeling room acoustics.

Overall, experimental results demonstrate the DARAS framework’s capability to accurately estimate room acoustic features and to approach actual measurement results, highlighting its significant potential and advantages in practical applications.
\vspace{-3mm}
\subsection{Ablation Study} \label{sec:Ablation Study}
To validate the effectiveness of the proposed DARAS model based on the MASS-BRPE module, we designed and conducted a series of ablation experiments in a real-world blind room scenario. The detailed results are shown in Table \ref{tab:AS}.

Firstly, to verify the effectiveness of the proposed hybrid-path cross-attention feature fusion method, we compared it with the naive feature-level fusion method. The naive feature-level fusion method concatenates room acoustic features $\bm{\mathbf{F}}_{\mathrm{a}}$ and audio features $\bm{\mathbf{F}}_{\mathrm{s}}$ directly along the feature dimension, followed by an MLP consisting of two linear layers for fusion. Additionally, experimental results demonstrate that even simple feature fusion methods can yield comparable performance to FiNS.

From these experimental results, it is evident that the proposed hybrid-path cross-attention fusion method outperforms the naive feature-level fusion method across multiple metrics such as $\mathcal{L}_\mathrm{STFT}$, ${\mathcal{L}}_\mathrm{MAG}$, $\mathcal{L}_\mathrm{SC}$, and MAE. Specifically, under identical Type conditions, performance improvements of approximately 11.6\% and 13.8\% were observed for $\mathcal{L}_\mathrm{STFT}$ and MAE metrics, respectively. These findings clearly indicate that the hybrid-path cross-attention fusion method effectively facilitates deep interactions between audio and room acoustic features, significantly improving blind room RIR estimation performance.

Secondly, we conducted detailed ablation experiments regarding the dimensions of room acoustic features input into the model. The specific experimental configurations are listed in Table \ref{tab:AS} (Types A-D), where $N_\mathcal{V}$ and $N_\mathcal{\zeta}$ represent dimensions of room acoustic features related to $\mathcal{V}$ and \rts, respectively.

By analyzing the experimental results from Type A to Type C configurations, we observed a consistent performance improvement in blind RIR estimation as the dimensions of the fused room acoustic features increased. This trend clearly suggests that appropriately incorporating room acoustic information helps capture more diverse features from real environments. Additionally, this fusion enhances the model's adaptability to the RIR estimation task, thus improving estimation accuracy.

However, in the Type D configuration, further increasing the dimensions of room acoustic features led to performance degradation. This indicates that simply increasing the dimensions of room acoustic features does not always benefit the model. Excessive emphasis on room acoustic features may result in information imbalance and redundancy, interfering with the model's ability to effectively learn critical features and thereby reducing overall performance.

\begin{figure}[!t]
\centering
\subfloat[]{\includegraphics[width=1.1in]{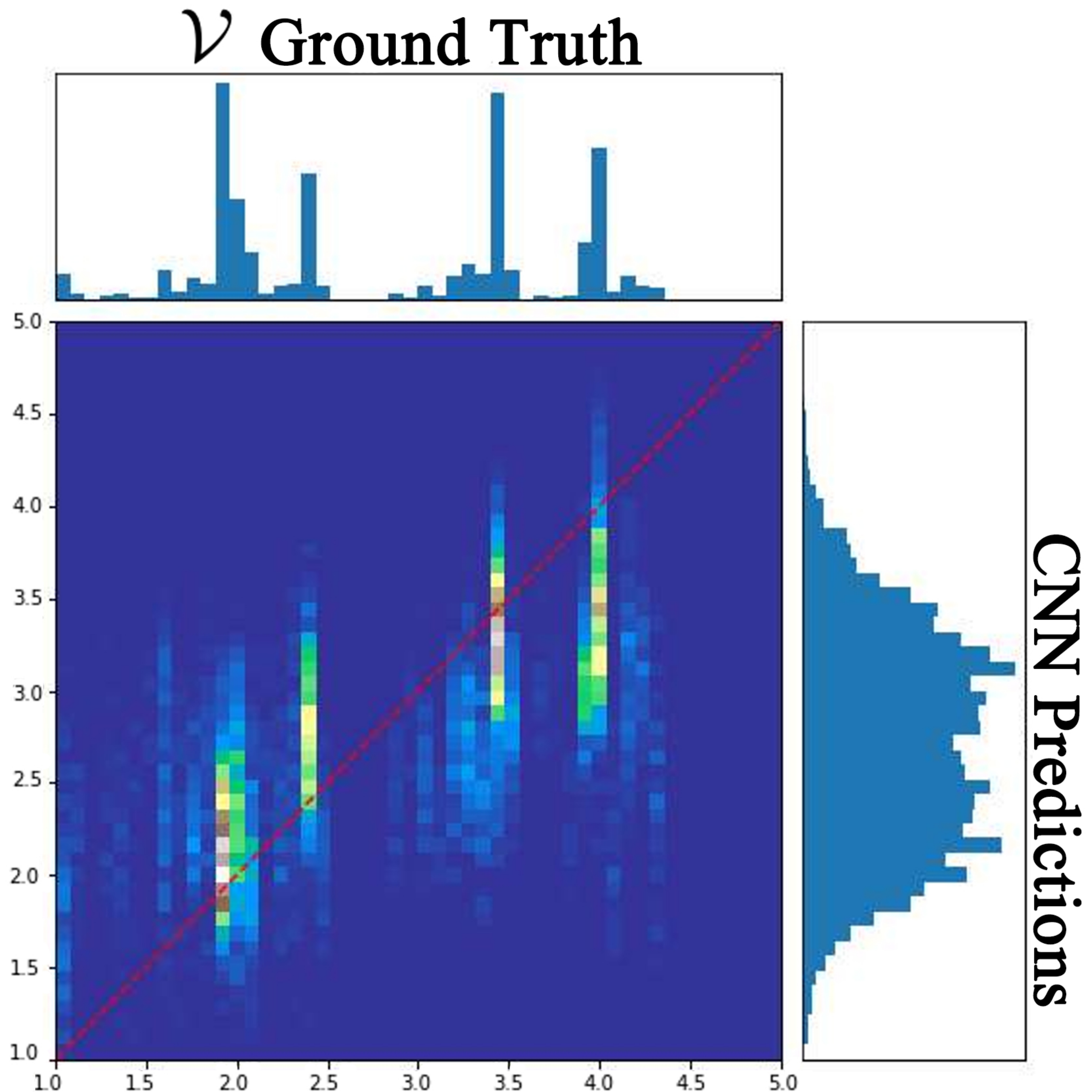}%
\label{CNN_V}}
\hfil
\subfloat[]{\includegraphics[width=1.1in]{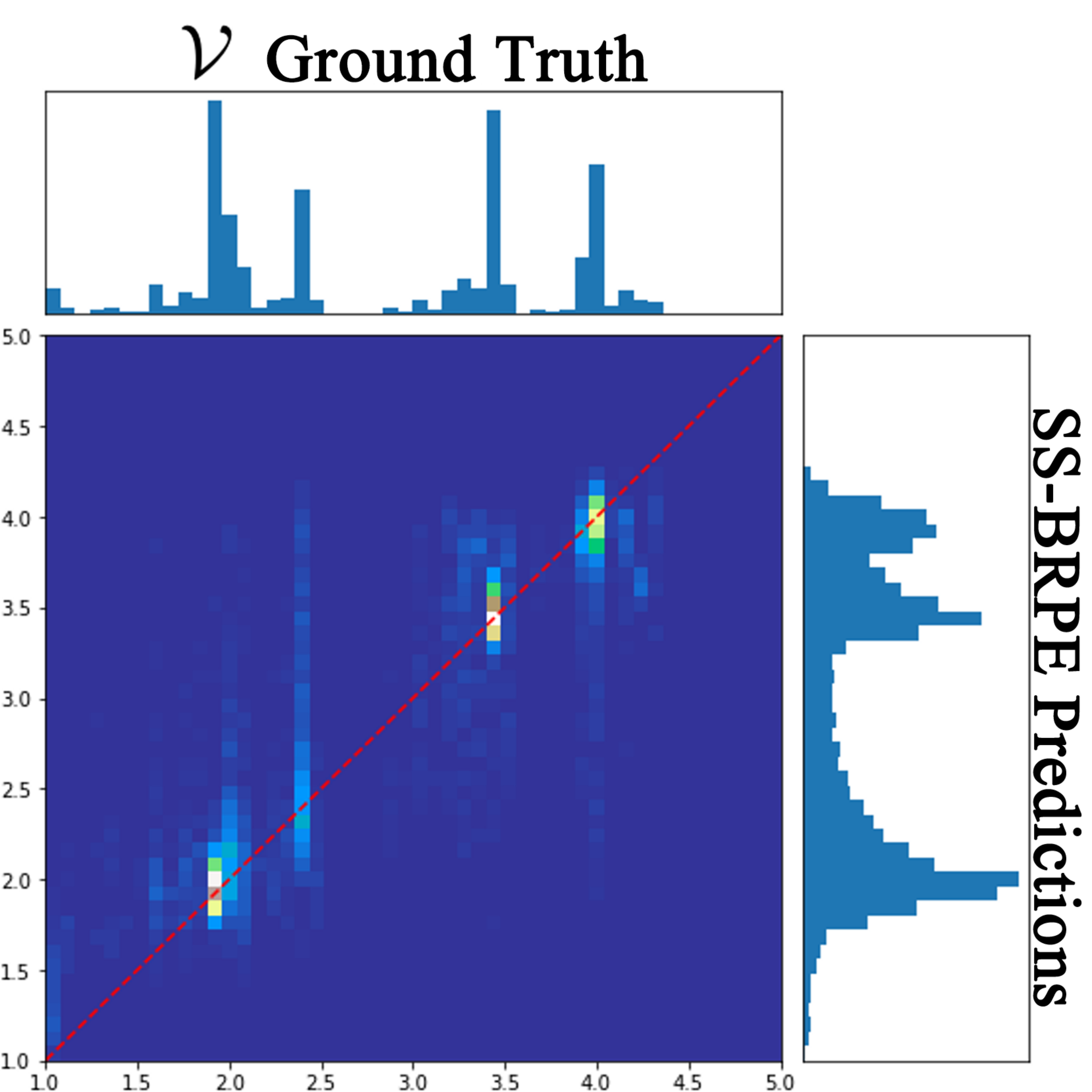}%
\label{SS_V}}
\hfil
\subfloat[]{\includegraphics[width=1.1in]{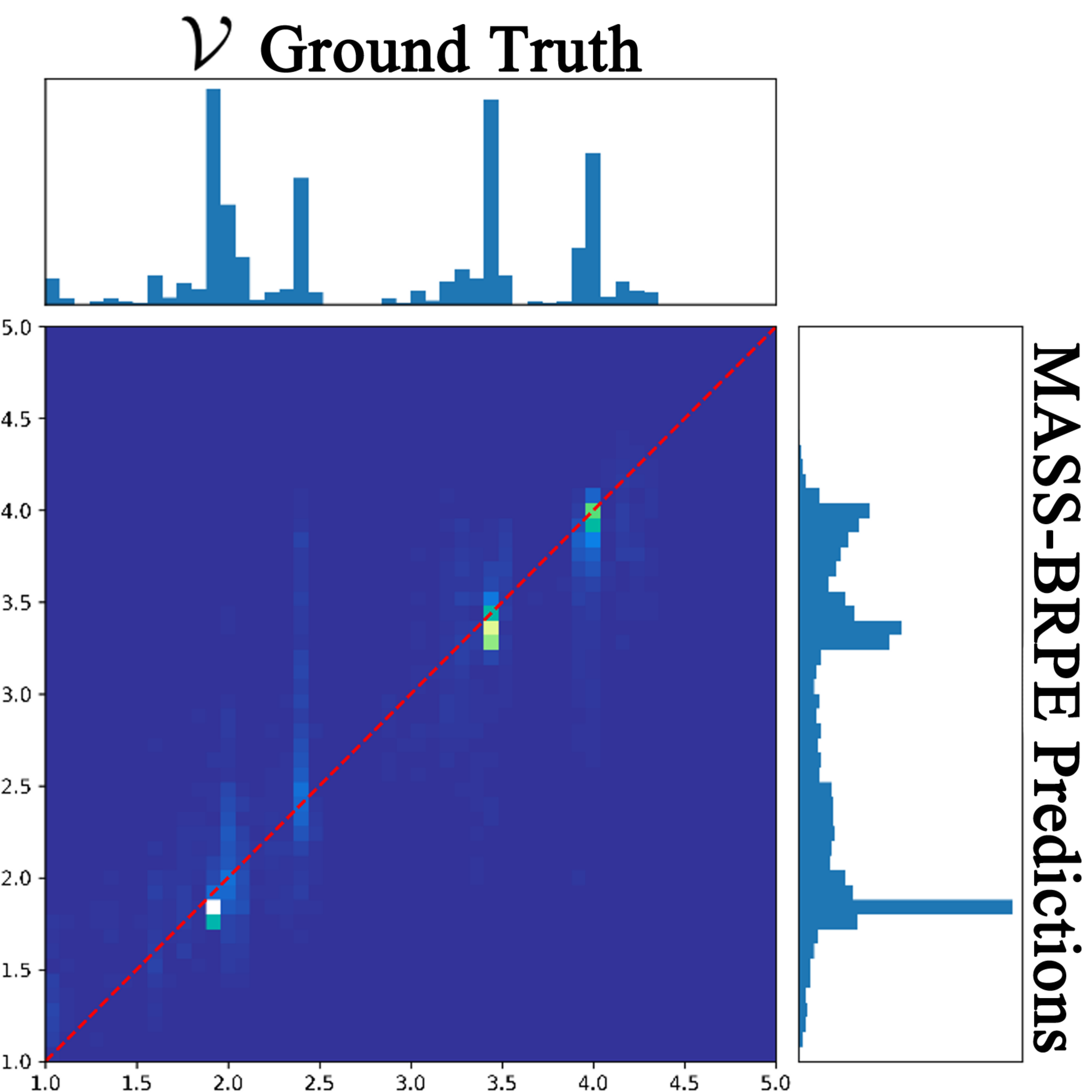}%
\label{MA_V}}
\vspace{-4mm}
\hfil
\linebreak
\subfloat[]{\includegraphics[width=1.1in]{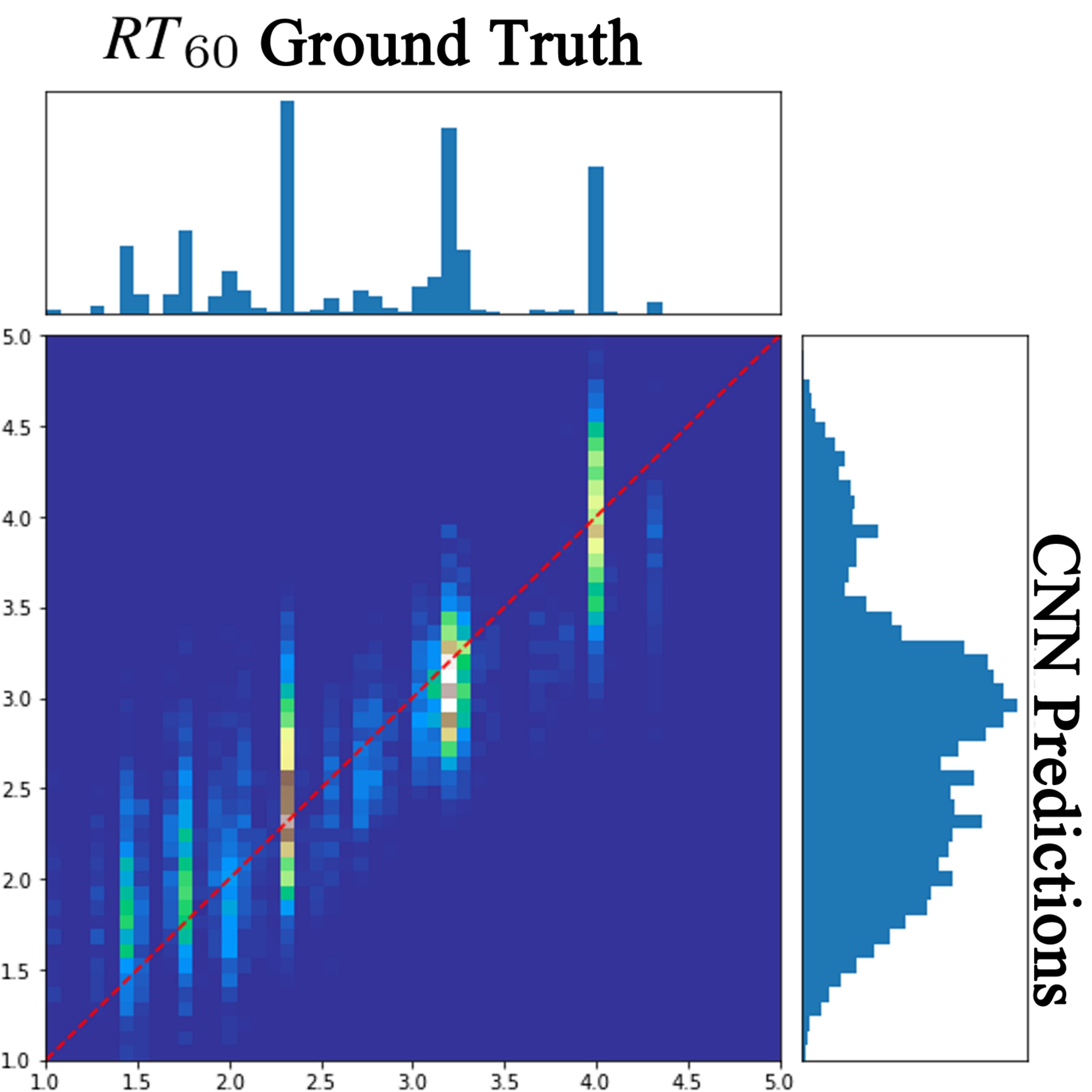}%
\label{CNN_RT}}
\hfil
\subfloat[]{\includegraphics[width=1.1in]{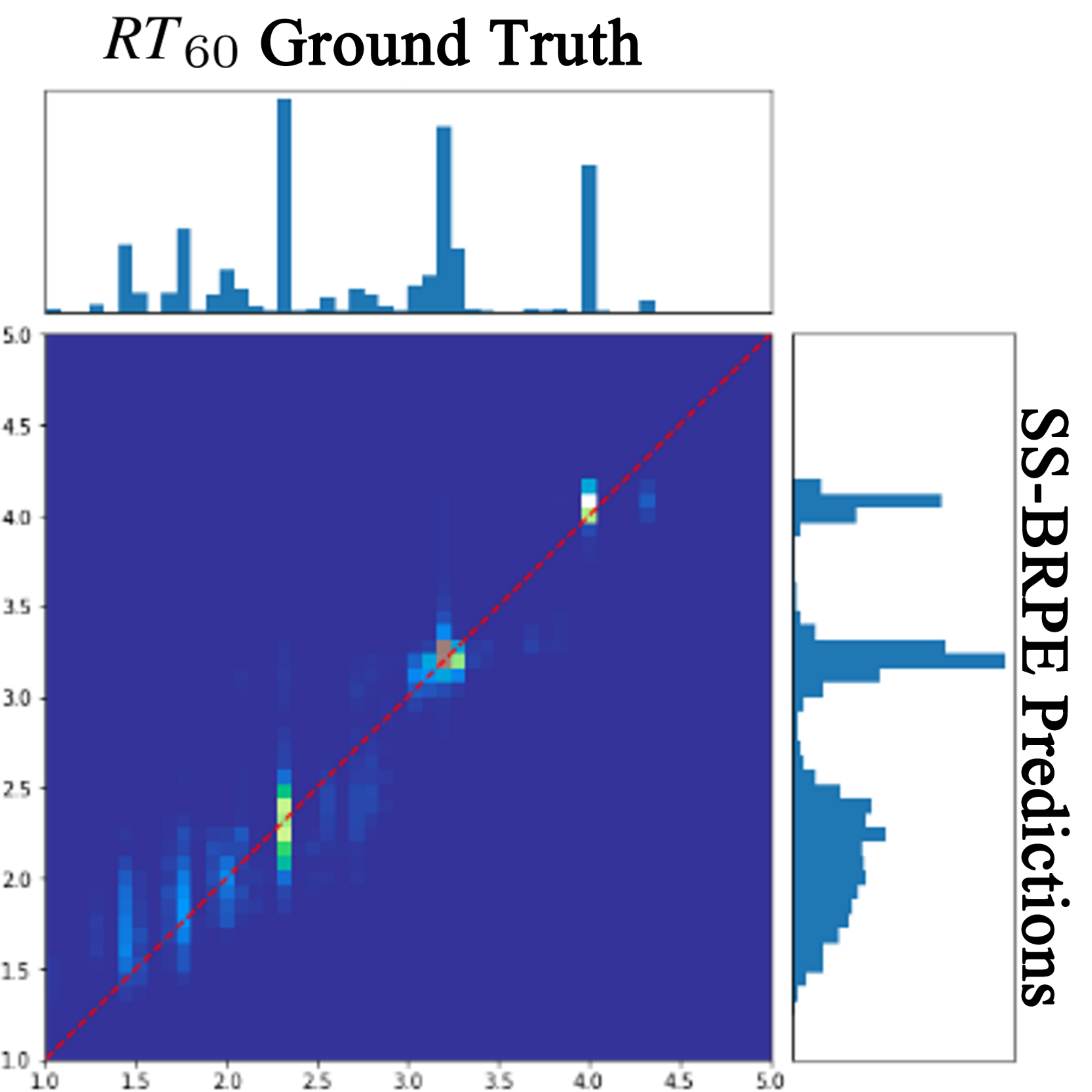}%
\label{SS_RT}}
\hfil
\subfloat[]{\includegraphics[width=1.1in]{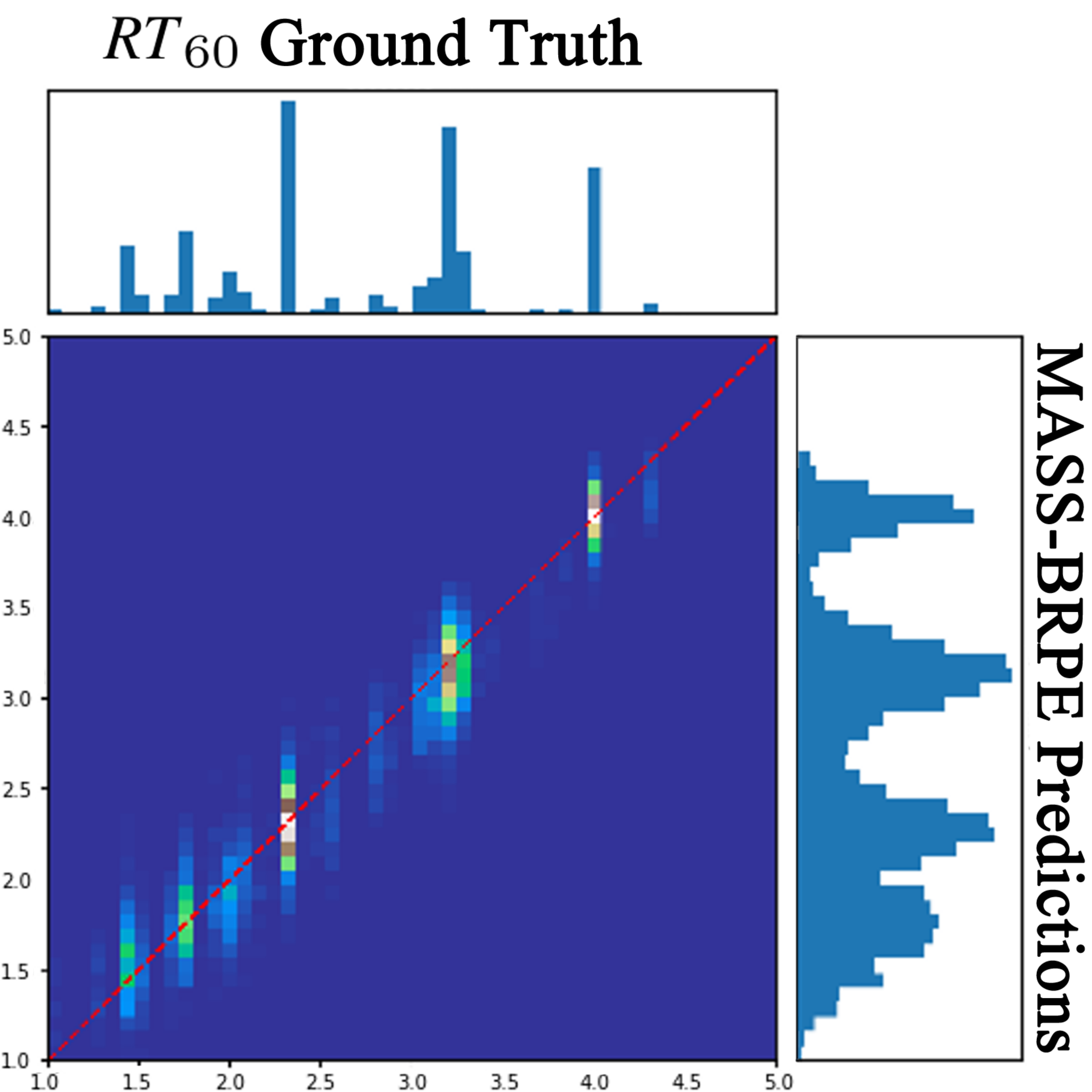}%
\label{MA_RT}}
\vspace{-4mm}
\hfil
\linebreak
\subfloat[]{\includegraphics[width=1.1in]{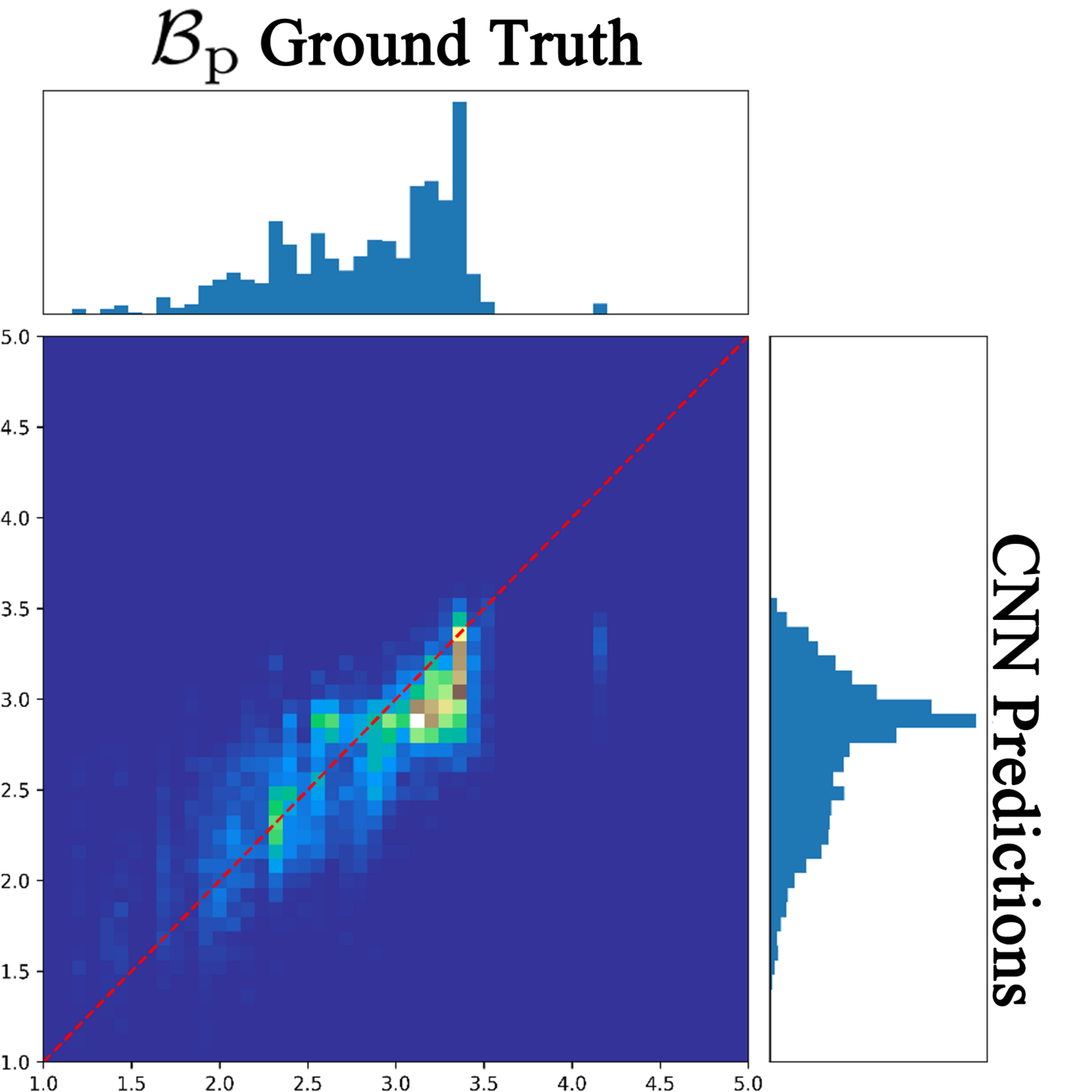}%
\label{CNN_BP}}
\hfil
\subfloat[]{\includegraphics[width=1.1in]{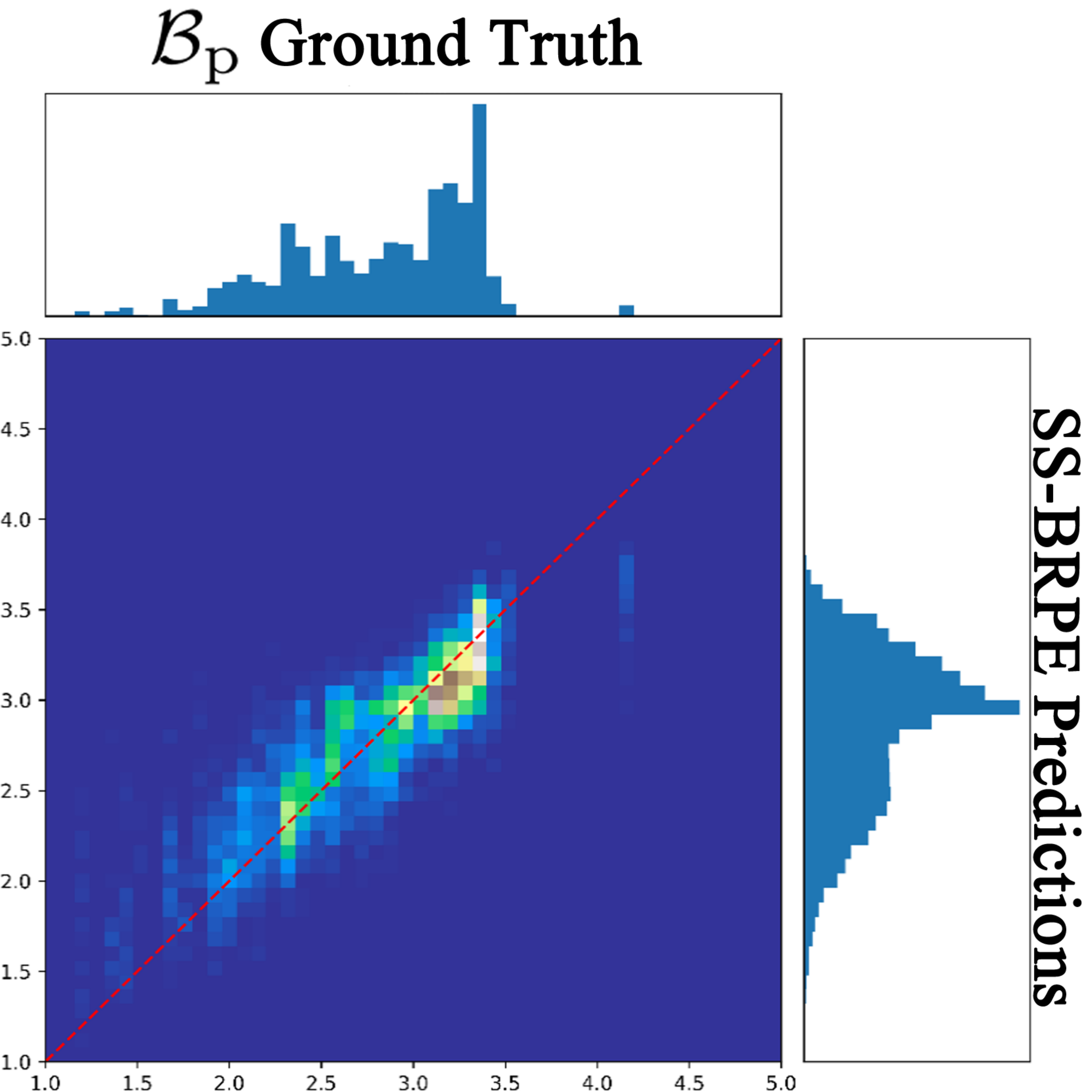}%
\label{SS_BP}}
\hfil
\subfloat[]{\includegraphics[width=1.1in]{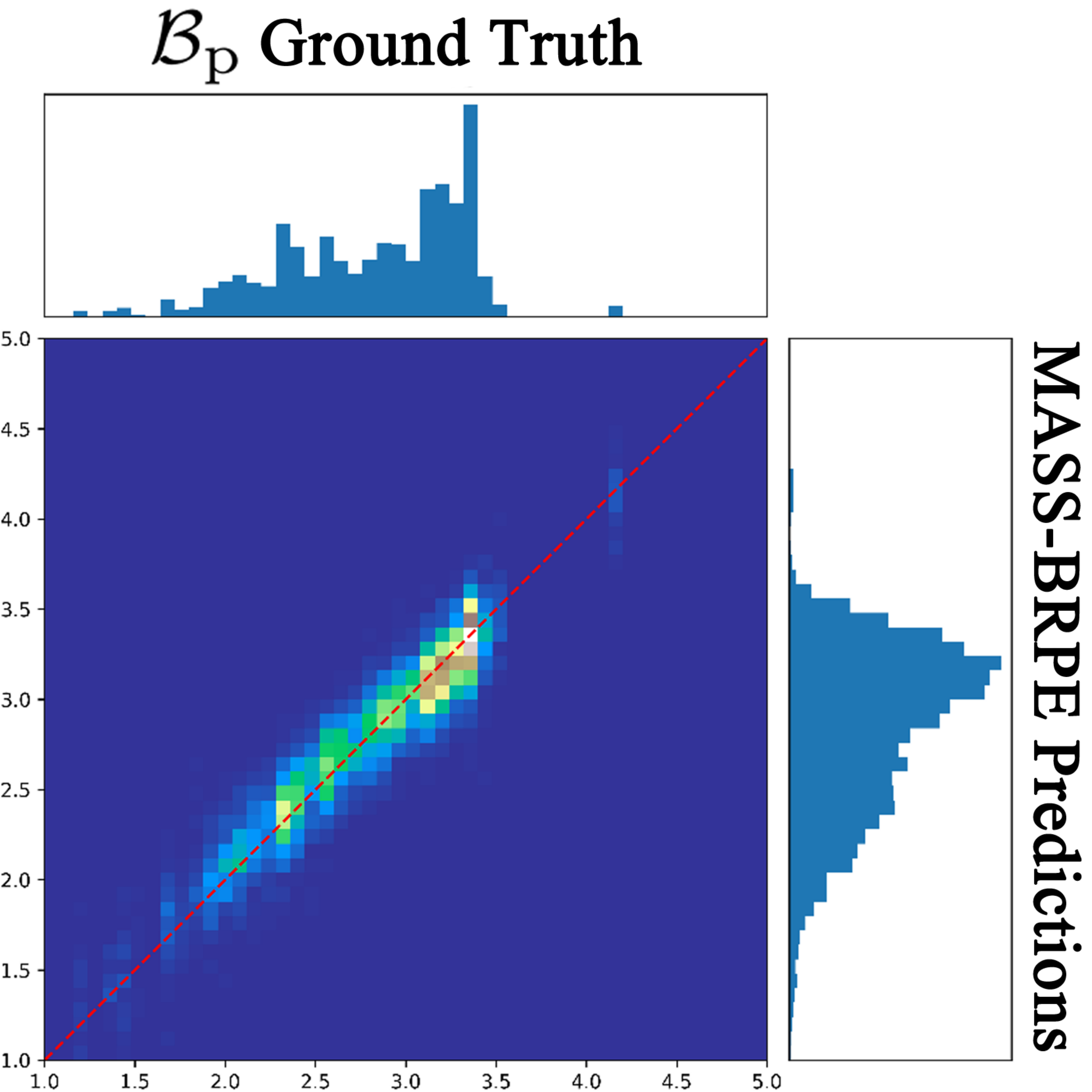}%
\label{MA_BP}}
\hfil
\vspace{-1mm}

\caption{Confusion Matrices of Blind Room Acoustic Parameter Estimation for Different Systems. (a) CNN-based module ($\mathcal{V}$).  (b) SS-BRPE module ($\mathcal{V}$). (c) MASS-BRPE module ($\mathcal{V}$). (d) CNN-based module (\rts).  (e) SS-BRPE module (\rts). (f) MASS-BRPE module (\rts). (g) CNN-based module ($\mathcal{B}_\mathrm{p}$, in samples, on a logarithmic scale). (h) SS-BRPE module ($\mathcal{B}_\mathrm{p}$). (i) MASS-BRPE module ($\mathcal{B}_\mathrm{p}$).}
\vspace{-6mm}
\label{fig: confusion matrices}
\end{figure}

\begin{figure*}[!t]
\centering
\captionsetup[subfigure]{skip=-5pt}
\subfloat[]{\hspace{-0.4cm}\includegraphics[width=1.4in]{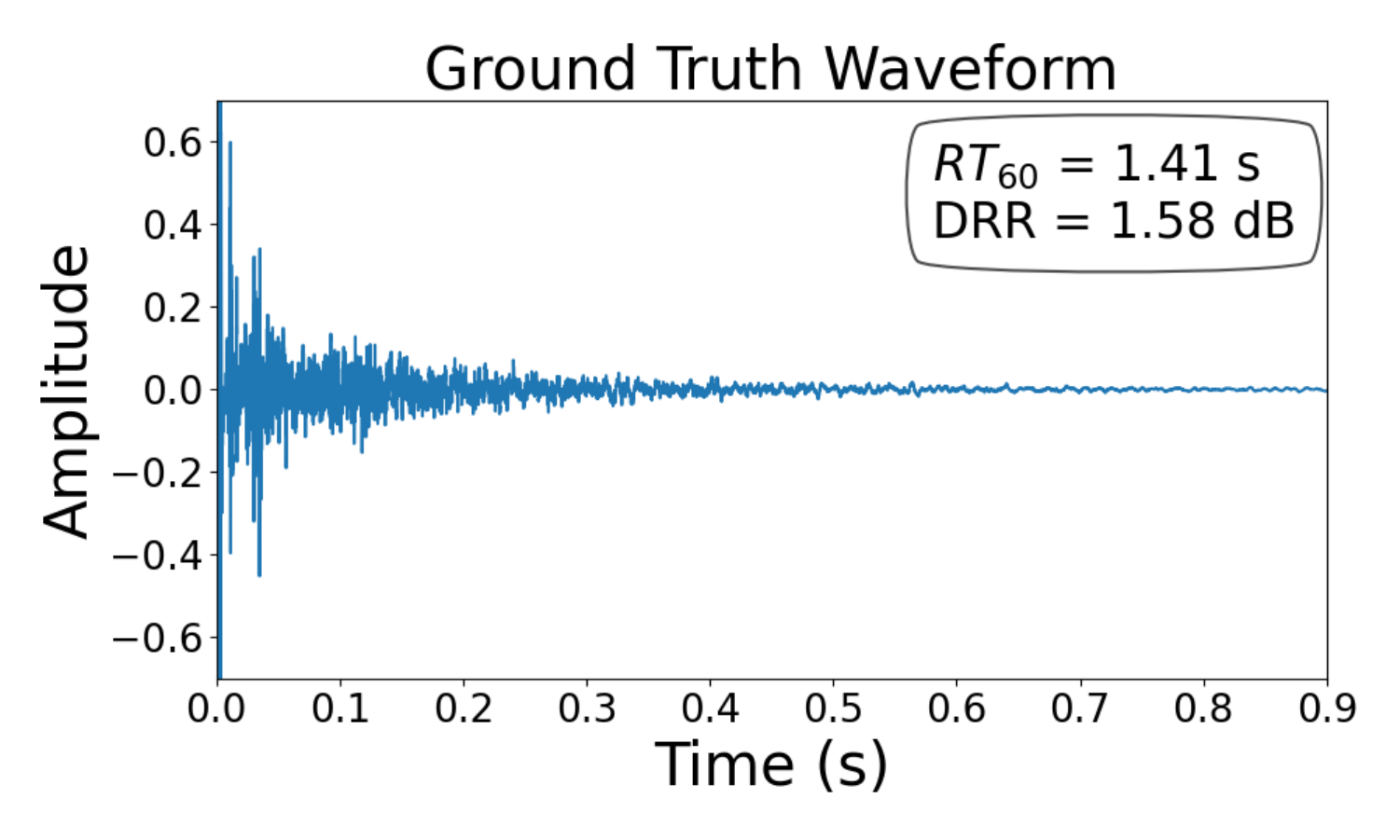}%
\label{TRUEWAVE}}
\hfil
\subfloat[]{\hspace{-0.58cm}\includegraphics[width=1.4in]{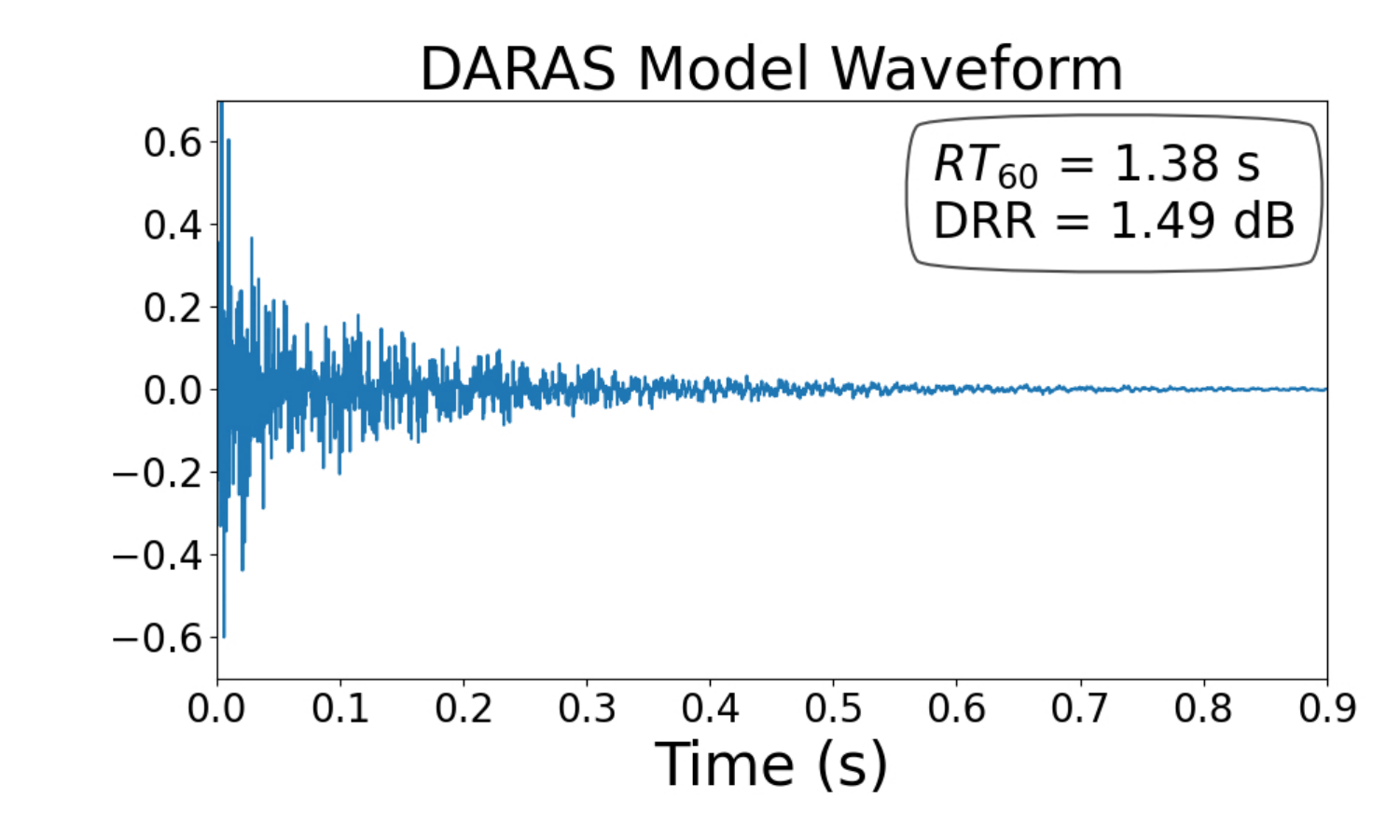}%
\label{PROWAVE}}
\hfil
\subfloat[]{\hspace{-0.58cm}\includegraphics[width=1.4in]{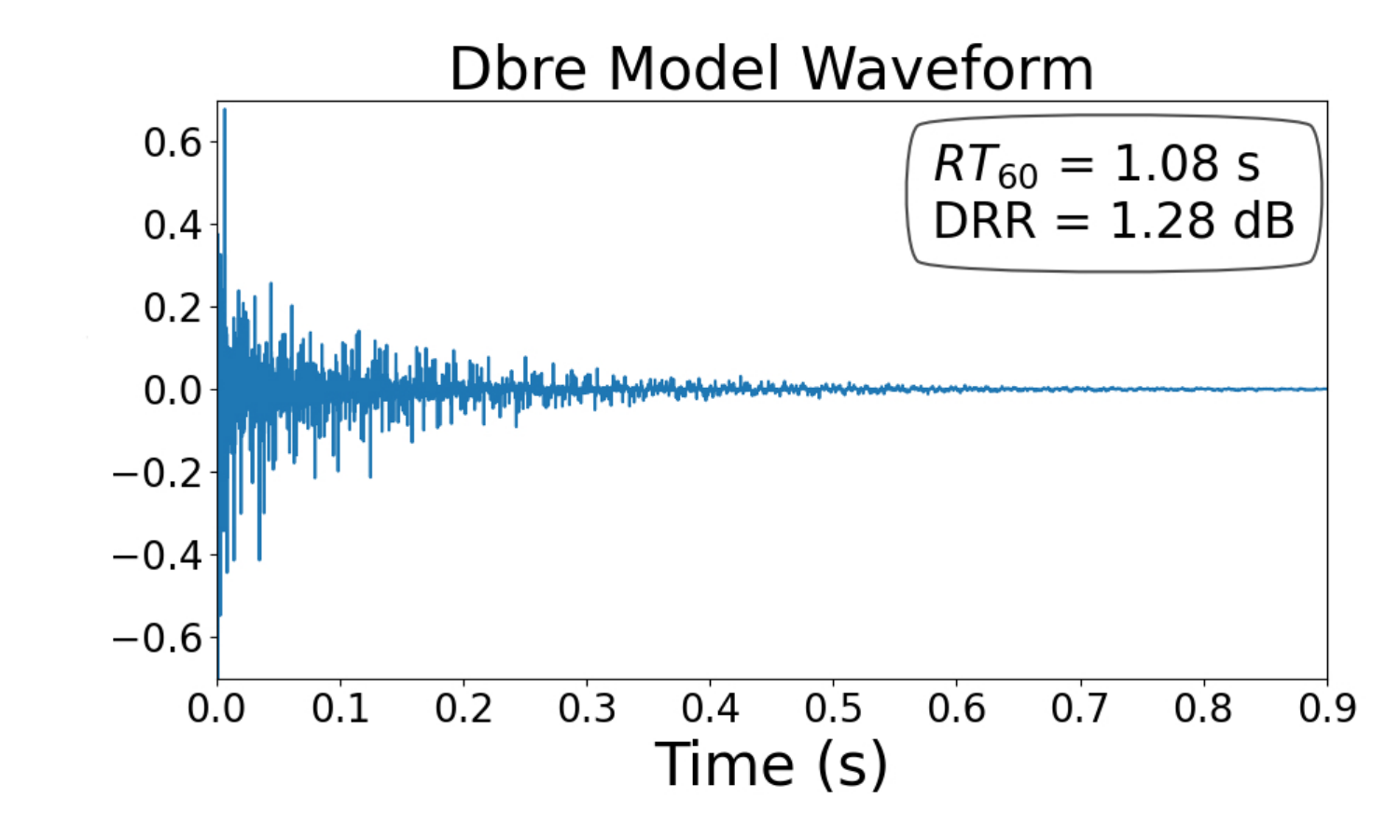}%
\label{DbreWAVE}}
\hfil
\subfloat[]{\hspace{-0.58cm}\includegraphics[width=1.4in]{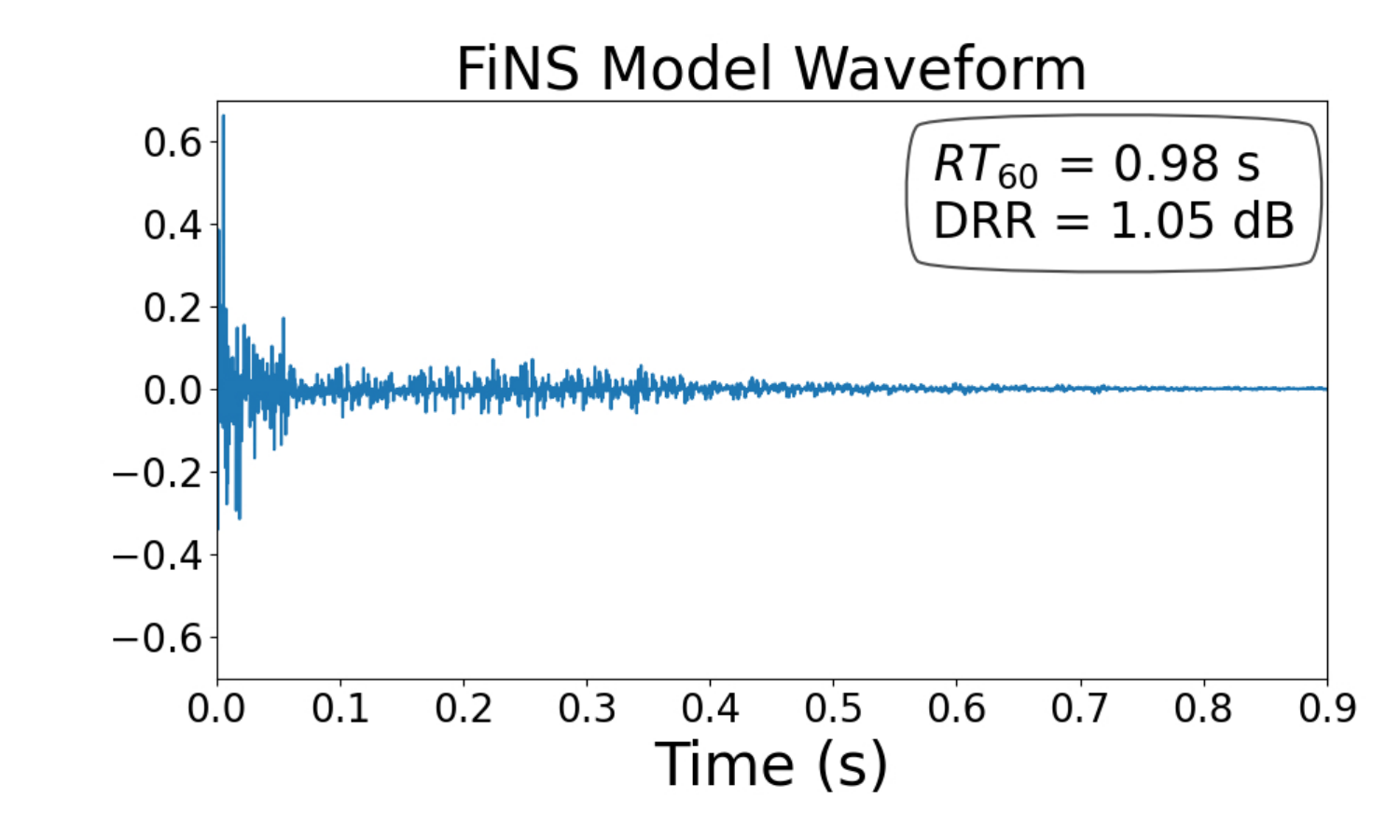}
\label{FINSWAVE}}
\hfil
\subfloat[]{\hspace{-0.58cm}\includegraphics[width=1.4in]{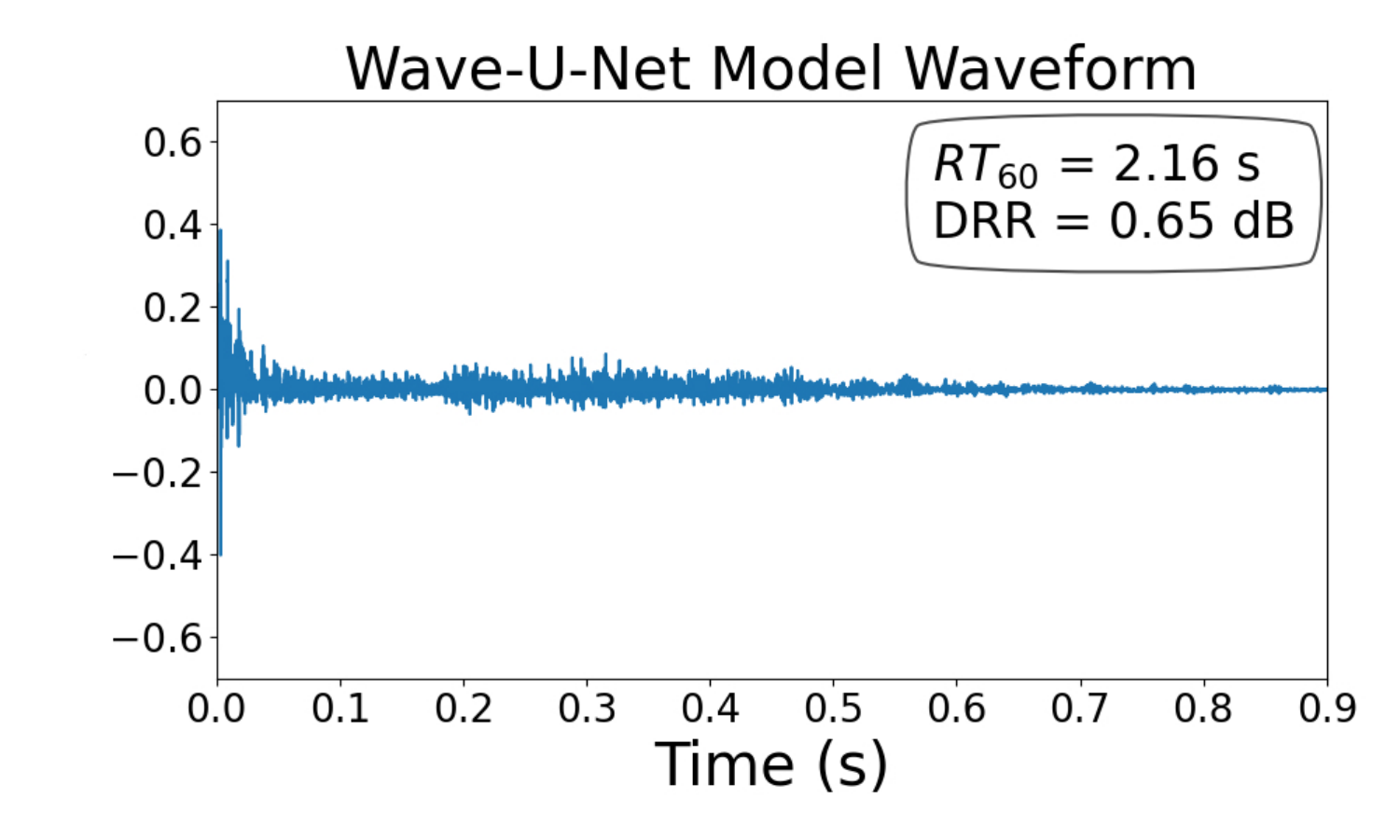}
\label{WAVWAVE}}
\vspace{-6mm}
\linebreak
\hfil
\begin{flushleft}
{\captionsetup[subfigure]{margin=48pt}%
\subfloat[]{\hspace{0.2cm}\includegraphics[width=1.2in]{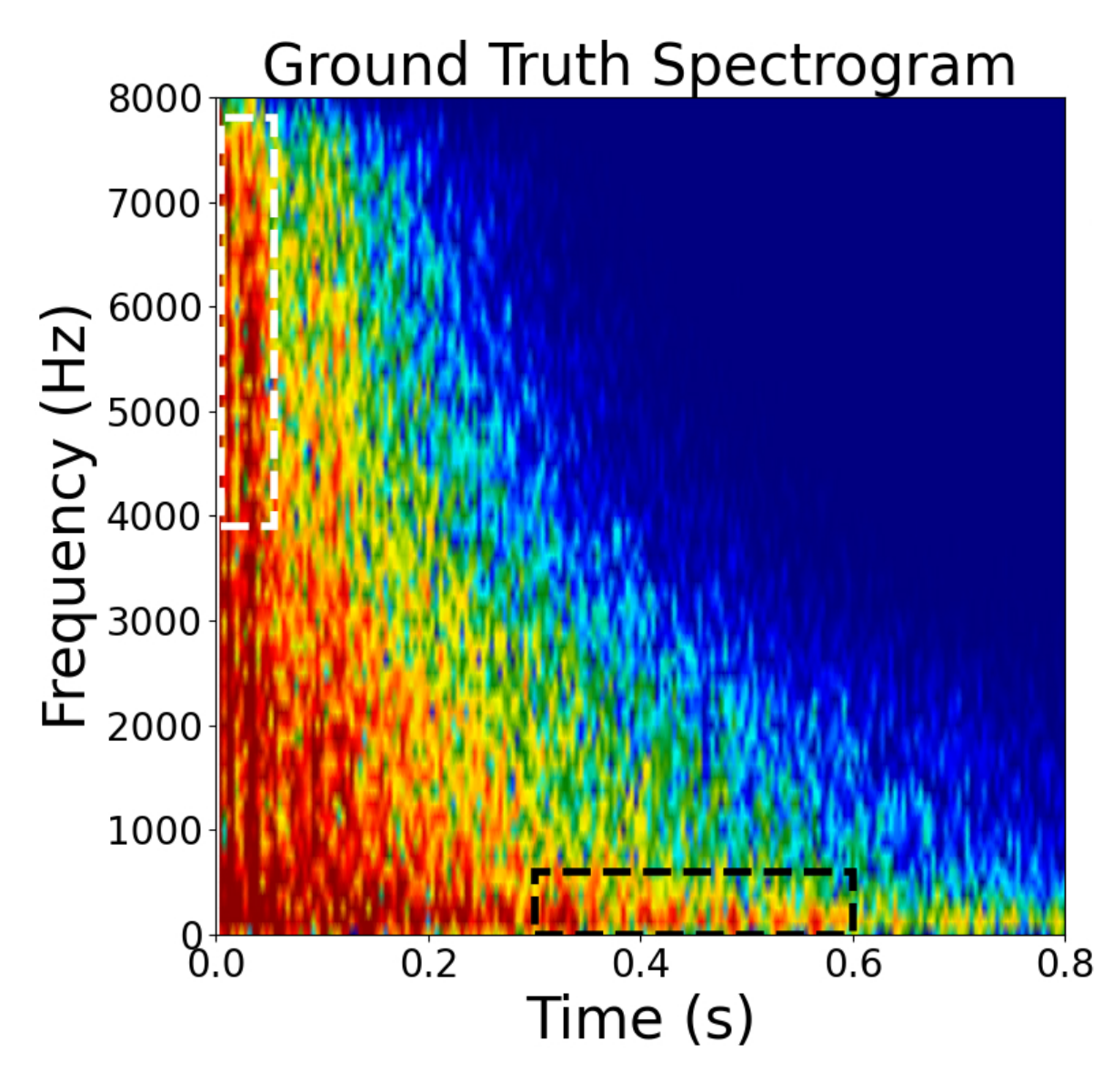}\label{TRUESPEC}}}
{\captionsetup[subfigure]{margin=52pt}%
\subfloat[]{\hspace{0.2cm}\includegraphics[width=1.2in]{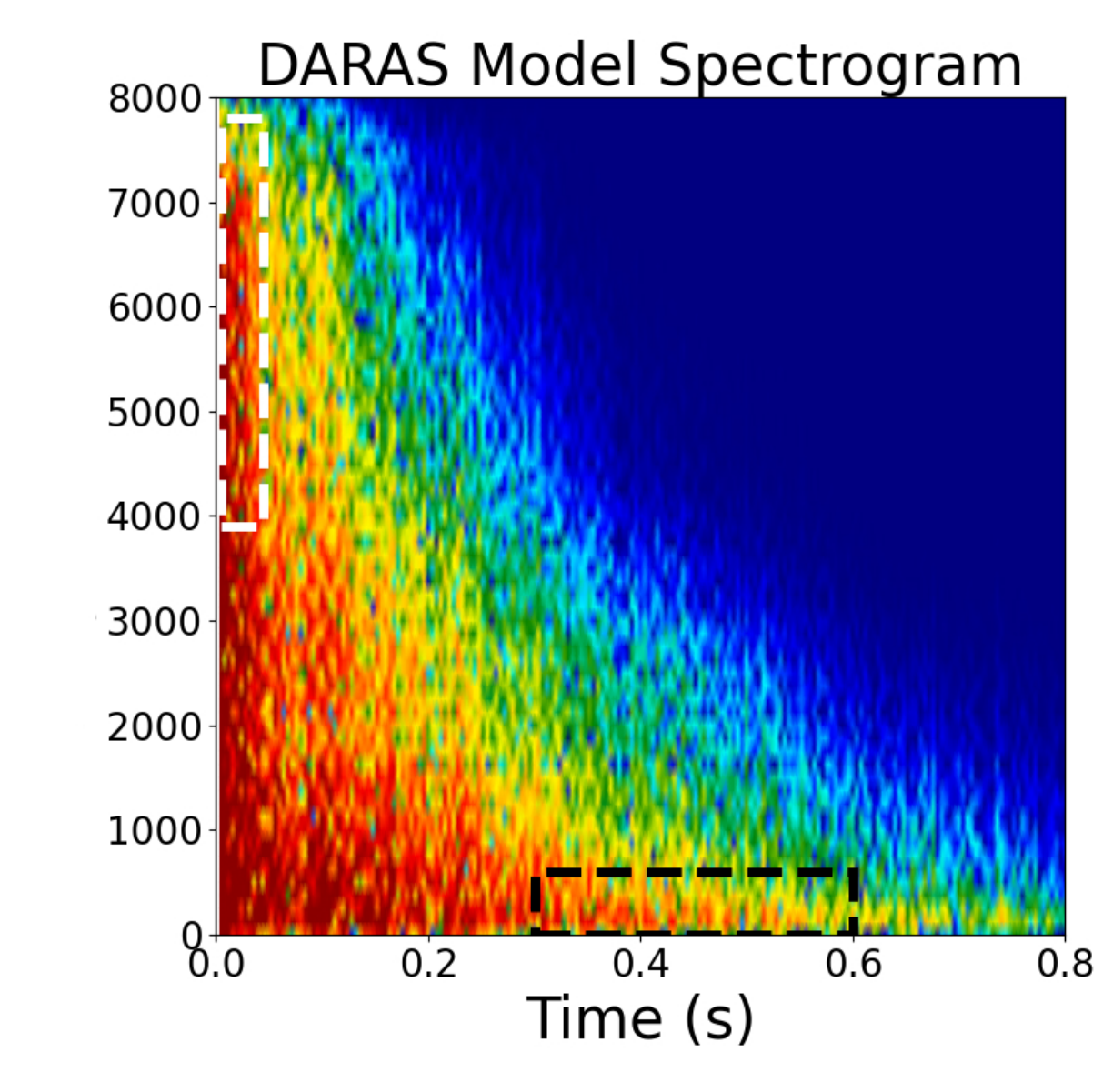}\label{PROSPEC}}}
{\captionsetup[subfigure]{margin=52pt}%
\subfloat[]{\hspace{0.2cm}\includegraphics[width=1.2in]{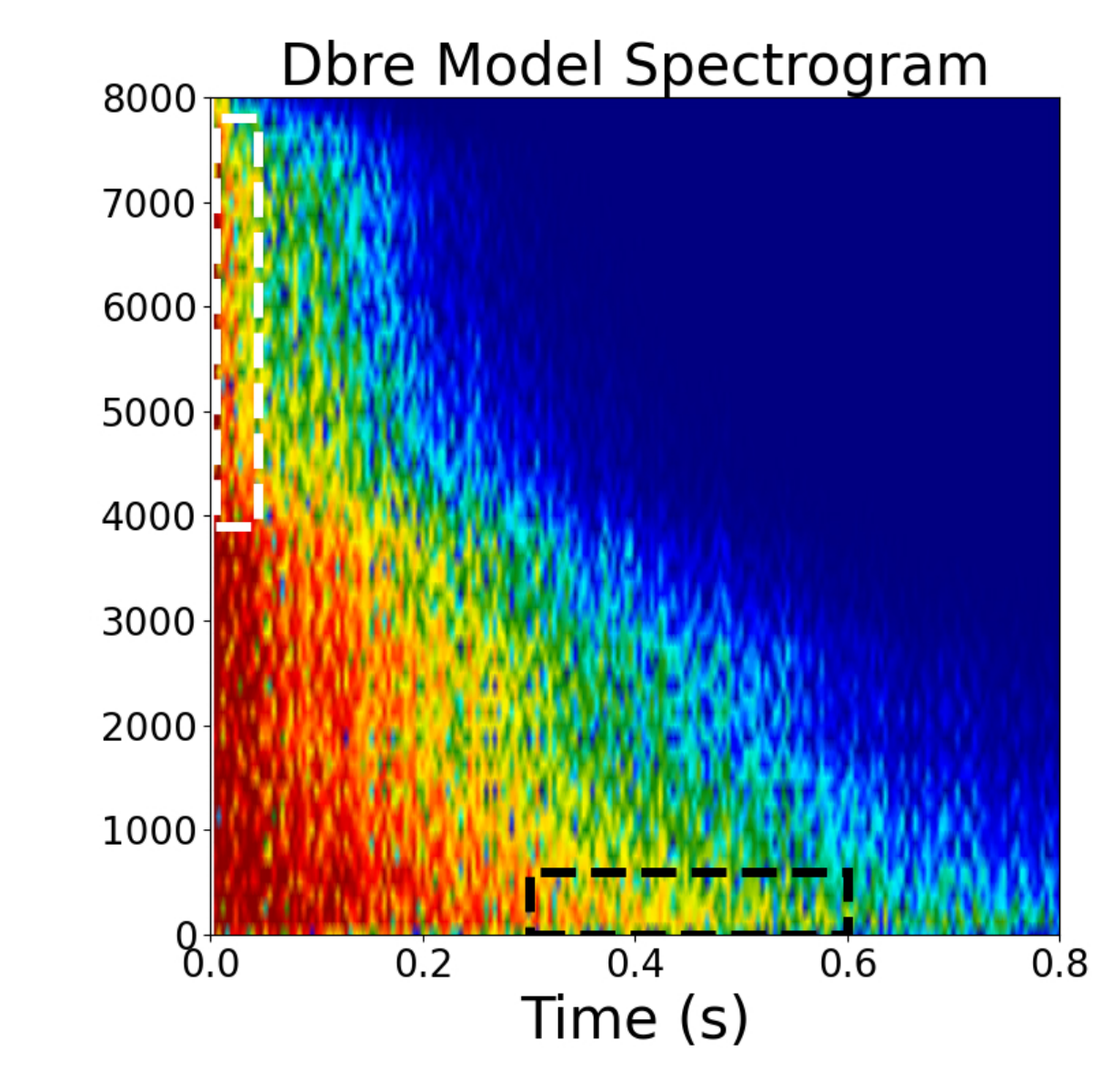}\label{DbreSPEC}}}
{\captionsetup[subfigure]{margin=53pt}%
\subfloat[]{\hspace{0.22cm}\includegraphics[width=1.2in]{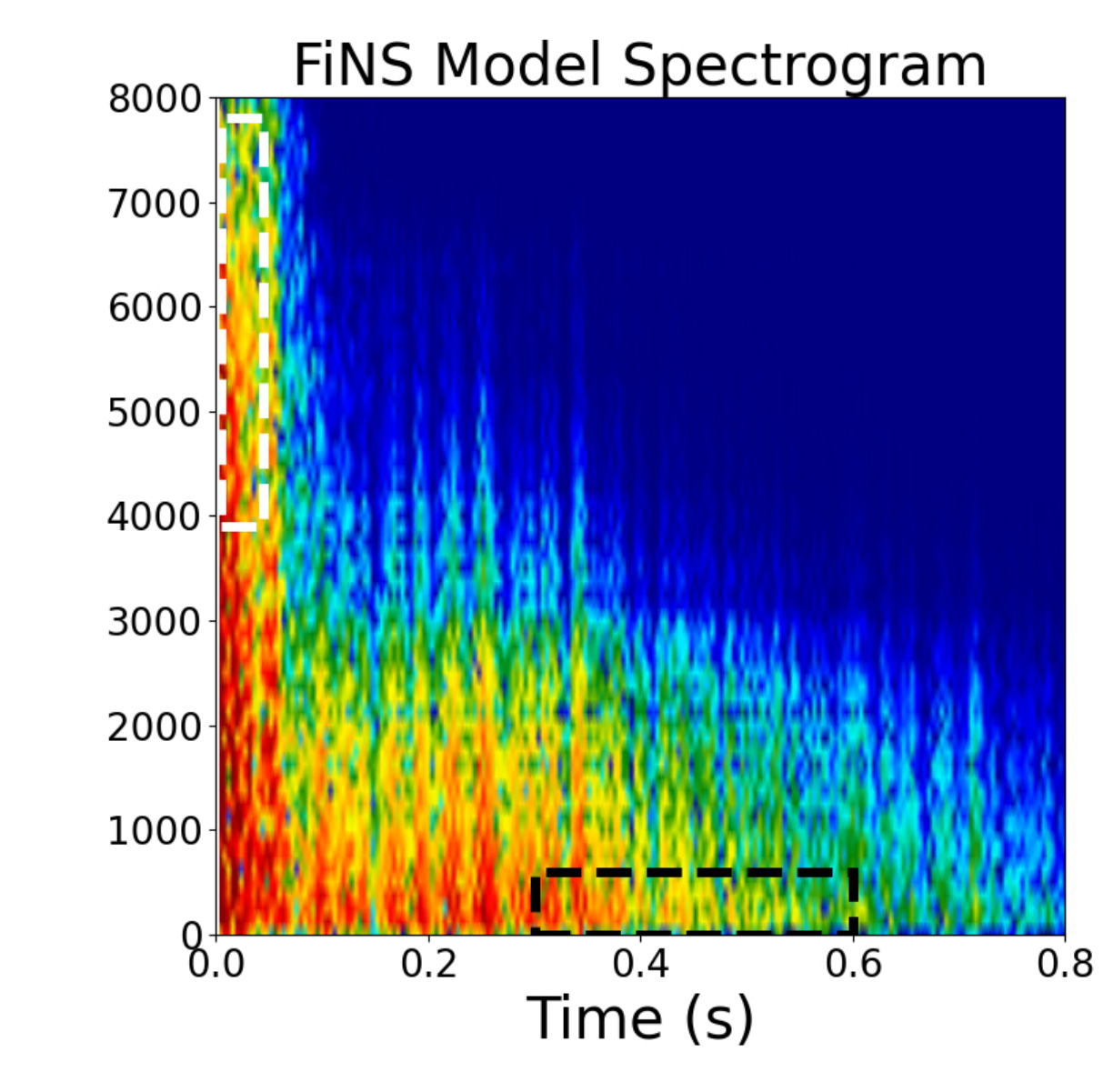}\label{FINSSPEC}}}
{\captionsetup[subfigure]{margin=56pt}%
\subfloat[]{\hspace{0.32cm}\includegraphics[width=1.2in]{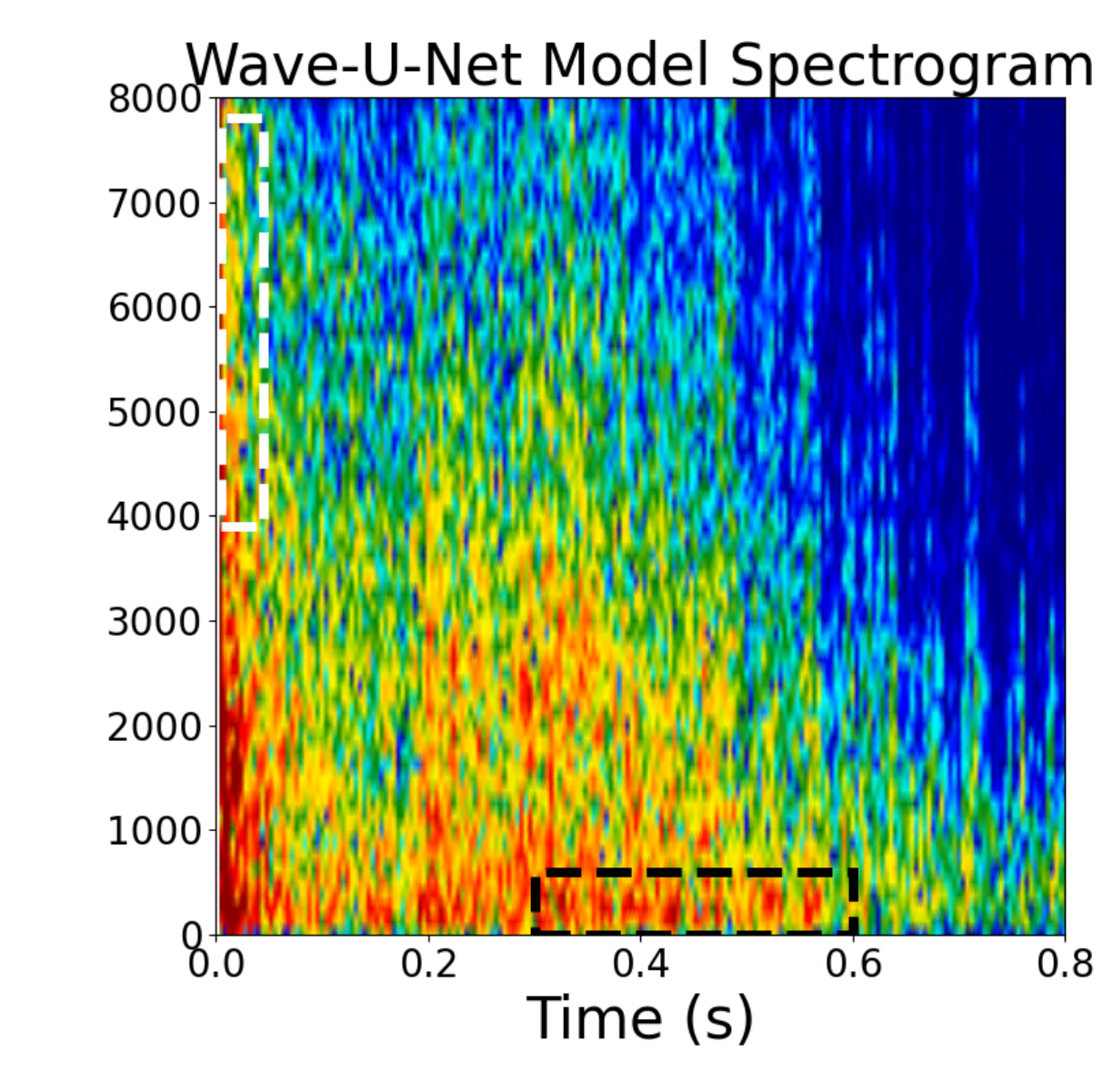}\label{WAVSPEC}}}
\subfloat{\raisebox{1.45ex}{\hspace{-0.12cm}\includegraphics[width=0.482in]{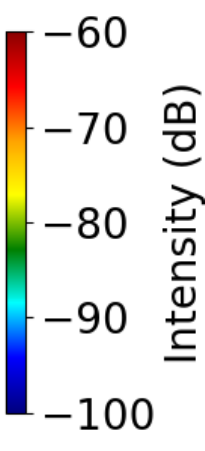}}\label{INT}}
\end{flushleft}
\vspace{-4mm}

\caption{Comparison of the RIR time-domain waveforms and spectrograms generated by different models. (a)–(e) show the time-domain waveforms for the ground truth, the proposed DARAS model based on the MASS-BRPE module, the Dbre model, the FiNS model, and the Wave-U-Net model respectively; (f)–(j) present the RIR spectrograms in the same order.}
\vspace{-5mm}
\label{fig: waveform and spec}
\end{figure*}

Furthermore, by comparing the model configurations of Type C with Type F and Type E with Type G, we explored the impact of the deep audio encoder on model performance.
In the absence of a deep audio encoder, we use a shallow time-domain encoder. The encoder’s convolutional block consists of a single 1-D convolution, followed by batch normalization and a PReLU activation. Residual connections are implemented with $1\times1$ convolutions, followed by batch normalization.
Results show that activating the deep audio encoder improved the $\mathcal{L}_\mathrm{STFT}$ and MAE metrics by approximately 4.7\% and 5.7\%, respectively. This demonstrates that a deeper audio encoder significantly enhances the model's feature extraction capability and overall performance. 
Furthermore, on a test set composed of unseen real-world rooms with diverse room types and extensive acoustic feature spans, the deep audio encoder consistently yields lower $\mathcal{L}_\mathrm{STFT}$ values across feature fusion methods and boundary point settings. This indicates stronger generalization and confirms its applicability to blind room RIR estimation.

Finally, to further investigate the influence of the boundary point dynamically set between early reflections and late reverberation, we conducted detailed comparative analyses between the Type C and Type E configurations, as well as between Type F and Type G. According to the results in Table \ref{tab:AS}, adopting a dynamic boundary point $\mathcal{B}_\mathrm{p}$, compared with a fixed boundary point at 50 $ms$, improved $\mathcal{L}_\mathrm{STFT}$ performance by approximately 6.1\%. Similar performance enhancements were also observed for the $\mathcal{L}_\mathrm{MAG}$, $\mathcal{L}_\mathrm{SC}$, and MAE metrics. 

\vspace{-5mm}
\subsection{Visualization}

We conducted a detailed visualization analysis to compare the performance of different modules in the BRPE task, aiming to gain deeper insights into each module's ability to capture room acoustic characteristics.

As depicted in Fig. \ref{fig: confusion matrices}, we present the confusion matrices for three distinct modules: the CNN-based module, the purely attention-based SS-BRPE module, and our proposed MASS-BRPE module. In these confusion matrices, the horizontal axis represents the true values of room parameters, while the vertical axis corresponds to the estimated values generated by each module. The evaluated parameters specifically included $\mathcal{V}$, \rts, and $\mathcal{B}_\mathrm{p}$. To clearly illustrate each model's performance, we applied a logarithmic scale to all parameters.

From the visualized confusion matrices, it is evident that our proposed MASS-BRPE module demonstrates outstanding fitting performance. The estimated values are densely concentrated around the true values, forming a clear diagonal trend. This indicates the high accuracy and consistency of our module in capturing room acoustic features. 


In summary, our experiments clearly demonstrate the superior performance of the proposed MASS-BRPE module in BRPE. This result highlights the module’s strong capability to extract high-level abstract representations that are closely related to room acoustics, thereby enabling more precise and robust blind RIR estimation.

Furthermore, we present examples of the estimated RIRs to demonstrate the outputs of the proposed DARAS model, the Dbre model, the FiNS model, and the Wave-U-Net model. We compared the RIRs generated by these four models with the real-world measured RIR in terms of time-domain waveforms and spectrograms. To ensure a fair comparison, we exclude BERP from this evaluation because it is not an end-to-end RIR generator. The ground truth is derived from actual measurements in an unseen lecture hall with a volume of approximately 370 $m^3$. As shown in Fig. \ref{fig: waveform and spec}, the time-domain waveform generated by the DARAS model aligns more closely with the actual measurement. The \rts\ and DRR estimated directly from the RIR also match the real values more accurately than those from the Dbre, FiNS and Wave-U-Net models. Meanwhile, the DARAS model provides more precise estimates of direct sound and early reverberation components, as is clearly evident in the frequency spectrum. As indicated by the white dashed box, the DARAS model demonstrates strong estimation accuracy in the high-frequency range of the direct sound. In addition, the DARAS model maintains a high degree of accuracy in estimating late reverberation, with a decay pattern closer to the real measurement. In comparison, the Dbre model shows a slightly faster late decay, the FiNS model exhibits biases in late-decay estimation, and Wave-U-Net struggles to reproduce noise-like late reverberation while also lacking precision in estimating early reverberation. Notably, the black dashed box highlights that in the low-frequency range of late reverberation, the DARAS model’s result is the closest to the actual RIR.

\vspace{-2mm}
\subsection{Subjective Listening Test}

To evaluate subjective perceptual differences of generated RIRs and their capability to reconstruct realistic reverberation scenarios, we conducted a MUSHRA\footnote{\href{https://github.com/audiolabs/webMUSHRA}{https://github.com/audiolabs/webMUSHRA}} test \cite{series2014method} to obtain subjective ratings of reverberation similarity.

In this listening experiment, we invited thirty self-reported normal-hearing listeners. Participants were instructed to rate the similarity between a reference reverberant audio signal and dry audio signals convolved with RIRs estimated by different models, including the proposed DARAS model, Dbre \cite{yapar2024demucs},  FiNS \cite{steinmetz2021filtered} and Wave-U-Net \cite{stoller2018wave}. The reference RIRs were obtained from real-world environment test sets. These audio samples encompassed diverse content, including one female and two male speech segments from the ACE dataset \cite{eaton2016estimation}, as well as movie clips and musical pieces in styles such as jazz and classical.
\begin{figure}[ht]
\centering
\vspace{-3mm}
\includegraphics[width=6cm]{Fig/MUSHRA.pdf}
\vspace{-3mm}
\caption{Results of the MUSHRA subjective evaluation comparing reverberation similarity ratings for different models, along with hidden reference and anchor conditions. *** indicates ${p}_{\text{adj}} \ll 0.0001$.}
\vspace{-3mm}
\label{fig:mushra}
\vspace{-1mm}
\end{figure}
Additionally, hidden anchors (audio generated using the average RIR from the training set) and hidden reference audio signals were included to ensure the reliability of the subjective evaluation. In total, six distinct 10-second audio samples were evaluated, each consisting of six stimuli.

Fig. \ref{fig:mushra} shows the results of the MUSHRA subjective evaluation, in which our proposed DARAS model obtained the highest ratings, significantly outperforming the other methods. Additionally, most participants were able to confidently identify the hidden reference signals, with limited variance observed in the scores for the reference system.


To quantitatively verify these observations, we performed a Kruskal–Wallis H-test on the aggregated subjective scores from all audio samples. The results revealed significant differences in the median ratings across groups (${H}_{\mathrm{kw}} = 885.71,\; {p}_{\mathrm{kw}} < 0.0001$). Subsequently, Conover’s post-hoc test with Holm correction was performed for all pairwise comparisons among the groups. The results showed that the differences between the hidden reference and the Dbre model ($p_{\mathrm{adj}} \ll 0.0001$), as well as between the DARAS model and the Dbre model ($p_{\mathrm{adj}} \ll 0.0001$), were both significant. Notably, among all pairwise comparisons, the $p_{\mathrm{adj}}$ value between the DARAS model and the hidden reference was the largest, and significantly higher than that for any other comparison. This indicates that, in terms of subjective evaluation, the proposed method produced RIRs most similar to the reference signals.


In summary, listeners generally identified RIRs generated by the DARAS model as samples with the highest similarity to reference signals, significantly outperforming the Dbre and FiNS models in subjective naturalness and authenticity. This indicates that the DARAS model can more accurately capture and reproduce the characteristics of real acoustic environments, resulting in generated RIRs that closely approximate real-world auditory scenarios.

\vspace{-3mm}
\section{CONCLUSION}\label{CONCLUSION}
\vspace{-2mm}

This paper proposes an innovative DARAS model that addresses existing challenges in blind RIR estimation, including difficulties in data collection, high annotation costs, and significant estimation errors in prior models. The DARAS model utilizes a deep audio encoder to extract nonlinear features from speech signals and incorporates a MASS-BRPE module to  estimate crucial room acoustic parameters and characteristics. Additionally, a hybrid-path cross-attention feature fusion module enhances deep interactions between audio and room acoustic features. Furthermore, a DAT decoder achieves adaptive segmentation and synthesis of early reflections as well as late reverberations, thereby enhancing the match between estimated RIRs and actual room environments.

Experimental results demonstrate that the DARAS model outperforms existing baseline methods in terms of accuracy and generalization. It estimates key acoustic parameters of real-world environments and generates RIRs. Subjective listening tests further indicate that listeners consistently rate the RIRs produced by the DARAS model as the most similar to those observed in real-world acoustic scenarios. These findings highlight the DARAS model as an effective solution for non-intrusive RIR estimation in real-world applications.

Future work may explore the impact of source and microphone responses \cite{srivastava2022realistic}, as well as the integration of visual modalities. Furthermore, the generalization ability of the DARAS model will be evaluated on larger real-world datasets.

\vspace{-4mm}



\bibliographystyle{IEEEtran}
\bibliography{refs}

\begin{IEEEbiography}[{\includegraphics[width=1in,height=1.25in,clip,keepaspectratio]{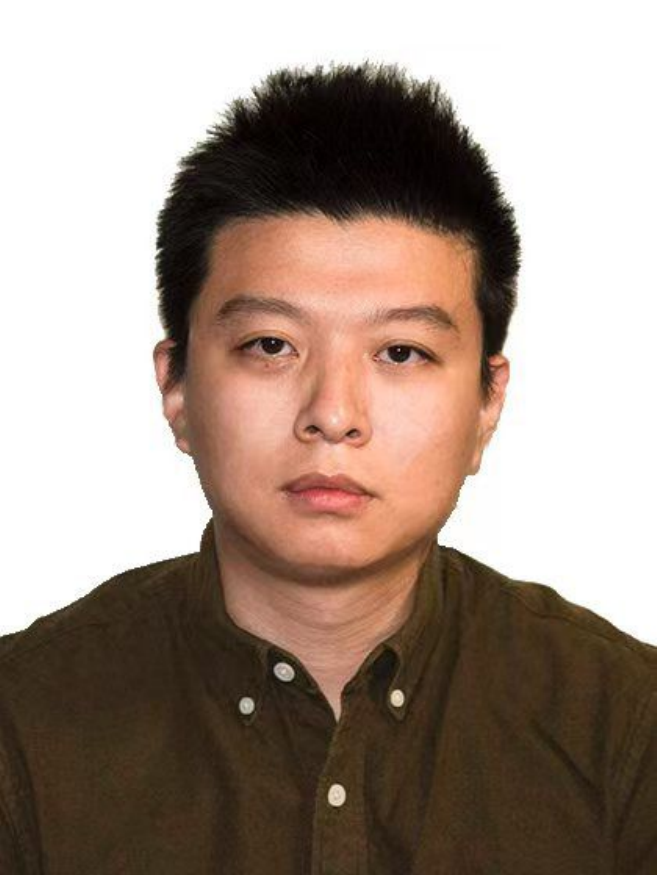}}]{Chunxi Wang}
(Student Member, IEEE) received the B.E. degree in communication engineering in 2022 from the Beijing University of Technology, Beijing, China, where he is currently working toward the Ph.D. degree in electronic science and technology.  His current research interests include room acoustics modeling and speech separation.
\end{IEEEbiography}

\vspace{-9mm} 

\begin{IEEEbiography}[{\includegraphics[width=1in,height=1.25in,clip,keepaspectratio]{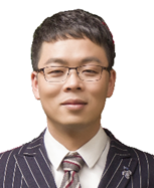}}]{Maoshen Jia}
(Senior Member, IEEE) received the B.E. degree in electronic information engineering from Hebei University, Baoding, China, in 2005, and the Ph.D. degree in electronic science and technology from the Beijing University of Technology, Beijing, China, in 2010. He is currently a Professor with the School of Information Science and Technology, Beijing University of Technology. His research interests include source localization, speech and audio coding, room parameter estimation, and array signal processing. He is currently an Associate Editor for the Circuits, Systems, and Signal Processing.
\end{IEEEbiography}

\vspace{-9mm} 

\begin{IEEEbiography}[{\includegraphics[width=1in,height=1.25in,clip,keepaspectratio]{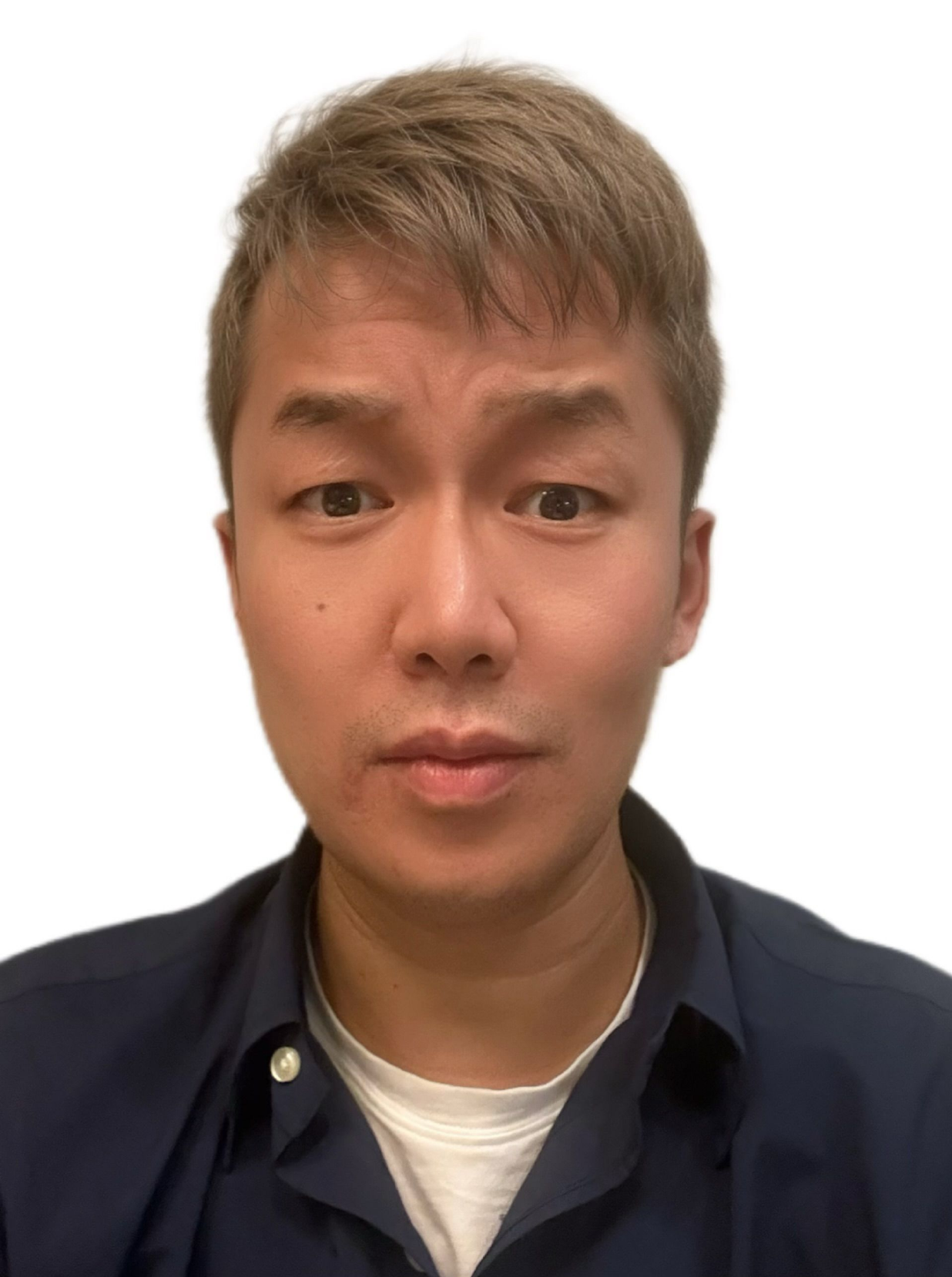}}]{Wenyu Jin}
(Member, IEEE) received the M.Sc. degree in system-on-chip from the University of Southampton, U.K., in 2010, and the Ph.D. degree in acoustic signal processing from Victoria University of Wellington, New Zealand, in 2015. He is currently a Co-Founder at Unseen AI Incorporated, Delaware, focusing on audio language models for real-world autonomy. Previously, he held positions as Engineering Manager at Cruise LLC, Principal Audio Research Engineer at Sonos, Inc.. His research interests include acoustic signal processing, audio/speech ML, spatial audio, and room acoustic modeling. He serves as Chair of the IEEE Signal Processing Society Twin Cities Chapter (2023–2027) and is a guest editor for EURASIP Journal on Audio, Speech, and Music Processing.
\end{IEEEbiography}


 




\vfill

\end{document}